\documentclass[prodmode]{acmlarge_tr}

\acmVolume{X}
\acmNumber{X}
\acmArticle{1}
\articleSeq{1}
\acmYear{2016}
\acmMonth{12}

\setcopyright{rightsretained}
\doi{0000001.0000001}

\issn{1234-56789}

\usepackage[ruled]{algorithm2e}
\usepackage{color}
\usepackage{tabularx}
\usepackage{graphbox}
\usepackage{amsmath}
\graphicspath{{.}}
\SetAlFnt{\algofont}
\SetAlCapFnt{\algofont}
\SetAlCapNameFnt{\algofont}
\SetAlCapHSkip{0pt}
\IncMargin{-\parindent}
\title{Improving Human-Machine Cooperative Visual Search With Soft Highlighting}
\author{RONALD T. KNEUSEL and MICHAEL C. MOZER \affil{University of Colorado}}
\begin{abstract}
Advances in machine learning have produced systems that attain human-level
performance on certain visual tasks, e.g., object identification. Nonetheless,
other tasks requiring visual expertise are unlikely to be entrusted to machines
for some time, e.g., satellite and medical imagery analysis. We describe a
human-machine cooperative approach to visual search, the aim of which is to
outperform either human or machine acting alone. The traditional route to
augmenting human performance with automatic classifiers is to draw boxes around
regions of an image deemed likely to contain a target. Human experts typically
reject this type of hard highlighting. We propose instead a soft highlighting
technique in which the saliency of regions of the visual field is modulated in
a graded fashion based on classifier confidence level. We report on experiments
with both synthetic and natural images showing that soft highlighting achieves
a performance synergy surpassing that attained by hard highlighting.

\end{abstract}


\acmformat{Ronald T. Kneusel and Michael C. Mozer. 2016. Improving Human-Machine Cooperative Visual Search With Soft Highlighting.} 

\begin{document}
\newcommand{\dprime}{{\ensuremath d'}}
\newcommand{\soc}[1]{{$\mathrm{SOC}_{#1}$}}

\begin{bottomstuff}
This research was supported by NSF grants SES-1461535, DRL-1631428, 
SBE-0542013, and SMA-1041755.  Authors' address:
R. T. Kneusel and Michael C. Mozer, Department of Computer Science, University 
of Colorado, Boulder, CO 80309-0430; email: ron@kneusel.org; mozer@colorado.edu.
\end{bottomstuff}

\maketitle

\section{Introduction}

In professions as diverse as diagnostic radiology, satellite imagery analysis,
and airport baggage screening, individuals are tasked with examining
complex visual images in search of target elements.  With sufficient training, 
individuals become skilled at such tasks, but expertise takes years to
acquire and human performance is fallible due to lapses of attention.  
To improve the accuracy and reliability of human performance,
\textit{computer-assisted detection (CAD)} systems have been developed that
digitally modulate images in order to increase the saliency of target elements.
Transformations such as histogram equalization, smoothing, sharpening, unsharp 
masking, and homomorphic filtering can guide viewer attention while preserving
critical image content \cite{Bankman2008,Kaur2011,Richards1999,Yasmin2012}.

Complementing these image-based transformations is the use of automatic pattern
recognition and classification techniques (hereafter, \textit{classifiers}) to
identify candidate target locations. Classifiers have a history in medical
imaging, where increased demands on radiologists necessitate means of
optimizing performance.  They are used both in CT and x-ray imagery and
commonly in mammography and lung-nodule detection.  The CAD system indicates
regions of interest (ROIs) in the image where a human expert should look,
either by outlining the regions with circles or boxes, pointing to the region
with an arrow, or drawing a symbol in the proximity of the region to indicate
the nature of the abnormality \cite{38,65,72,118,162,184,185}.  The form of
these \textit{highlights} has not changed significantly in the past twenty 
years.  They are in widespread use today and US federal health programs
reward physicians for using CAD. Although their efficacy has been 
demonstrated in laboratory studies \cite{5,18,65,181},
clinical studies are often discouraging \cite{162}, and
a recent large-scale study of practicing radiologists found that CAD does
not improve diagnostic accuracy of mammography on any metric assessed
\cite{Lehman2016}.

Highlighting is also common in remote-sensing applications involving aerial and
satellite-based imagery of the earth. To assist human analysts, classifiers
assign a class label to each pixel of the image conditioned on the local
spatial patch centered on the pixel. These class labels indicate geographic
features (e.g., mountains versus rivers versus land, agricultural versus urban
versus undeveloped areas) and are represented to the human via \textit{thematic
maps} in which false pixel coloring indicates the most likely class
\cite{8,214}.  The thematic map is often presented side-by-side with the
original image. Various techniques have been developed to indicate classifier
uncertainty, including: the use of whitening (interpolating toward white)
\cite{84} or manipulation of saturation \cite{128} to represent confidence,
the use of a display toggle to alternate between the thematic map and a
confidence map \cite{213}, and stochastic animations in which classifier
probabilities dynamically resample labelings \cite{22}.

There appears to be a dissociation in the CAD literature between (1) tasks
involving search for one or a small number of visual elements in an image,
typically medical images, and (2) tasks involving characterization of spatial
topography with multiclass labeling of individual pixels, typically
geospatial-sensing images. In the latter tasks, researchers have investigated
whether human analysts benefit from interfaces that encode classifier
uncertainty, yet in the former tasks, we are not aware of existing software
tools that provide an explicit representation of uncertainty.

\begin{figure}[tbp]
\centering
\includegraphics[width=137mm]{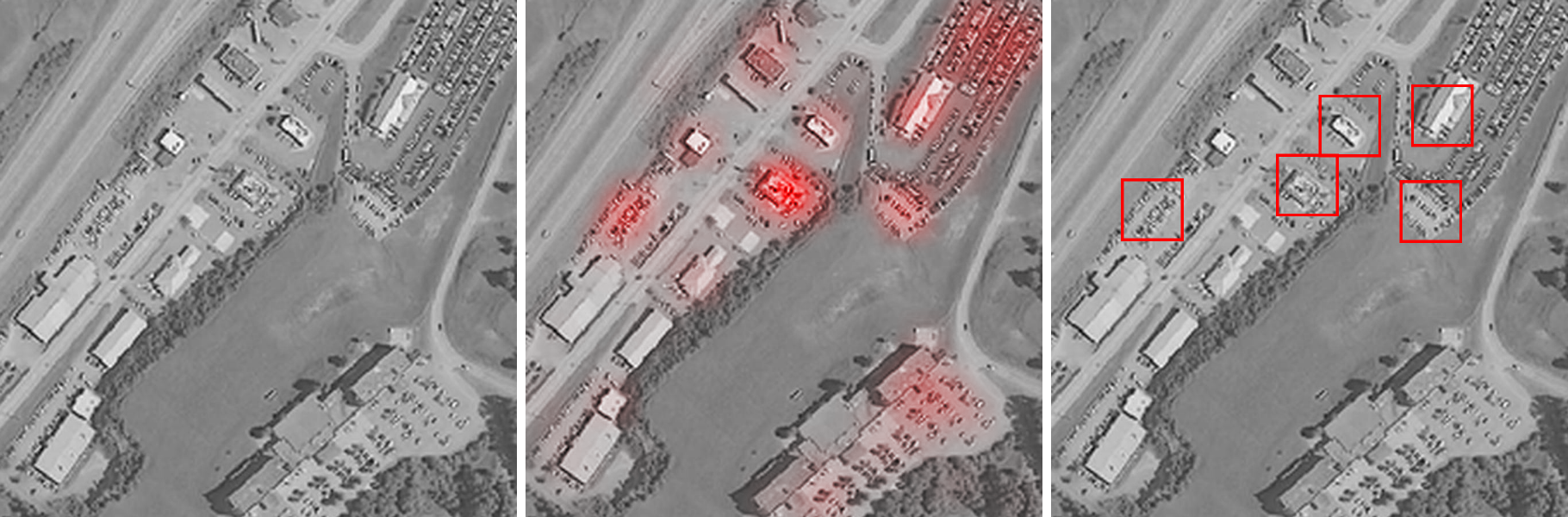}
\caption{(left) satellite image of an urban area; (middle) 
soft highlighting, where the saturation indicates the likelihood of
a target; (right) hard highlighting, where a box is drawn around
regions highly likely to contain a target}
\label{fig:hardversussoft}
\end{figure}

In this article, we
conduct studies to determine whether observers benefit from visualizations of
classifier uncertainty in \textit{search} tasks. We contrast the \textit{hard}
highlighting that has been used to indicate candidate targets---boxes enclosing
and arrows pointed at regions of interest---with \textit{soft} highlighting
that indicates not only target locations but classifier confidence.
Figure~\ref{fig:hardversussoft} illustrates these two types of highlights. The
soft highlights provide a graded saliency cue related to classifier confidence.
We are aware of only one other study that contrasts soft
and hard highlights: the
%---{\em analog} and {\em binary} in their jargon: the
very recent work of \citeN{CunninghamDrewWolfe2016}. We return to this 
research in the General Discussion.

For visual search and identification tasks, we hypothesize that 
CAD systems can be significantly improved by incorporating soft highlighting 
over hard highlighting. To make this argument, we briefly discuss issues
arising with hard highlighting.

\subsection{Hard highlighting: There is no free lunch}
\label{sec:no_free_lunch}

In \citeN{Krupinski_111}, radiology residents briefly inspected images in which
circles were used to highlight an ROI in chest x-rays.  The presence of
highlights improved discrimination of tumors from non-tumors in the ROI (see
also \citeN{65}).  However, the highlights yielded poorer detection of tumors
outside the ROI.  In this study, targets appeared in the highlighted region on
half of all trials.  Consequently, the highlight provided no useful information
about target presence. Thus, the study served to evaluate the effect of
highlighting on the allocation of attention.

\citeN{Zheng2001} tested board-certified radiologists and manipulated
true-positive (TP) and false-positive (FP) rates of the classifier. With a
highly accurate classifier---a TP rate of 0.9 and a mean of 0.5 FP highlights
per image---highlights improved discrimination performance. However, with
relatively poor classifiers---having either a lower TP rate or a higher FP
rate---highlights decreased performance, mainly in terms of increased failure
to detect non-highlighted targets.  % masses and microcalcifications

\citeN{Zheng2004} found little effect of highlighting if the highlight onset
follows an initial interpretation by the radiologist. However, if the
highlights appear simultaneously with the image, performance is impaired, both
in a detection task and in a benign/malignant classification task. As in
earlier studies, performance deficits manifest in terms of a lower rate of
detecting nonhighlighted targets. Further evidence that FP highlights
misdirect individuals away from discovering an unmarked target,
\citeN{5} discovered that highlighting the wrong location resulted
in twice as many missed targets as failing to highlight any location.

Mirroring studies conducted with medical personnel on medical imagery,
\citeN{Drew2012} performed a traditional laboratory search task in
which participants located a target letter T among L's in displays with
cloud-like noise whose spatial-frequency distribution matched that of
mammographic images. A no-highlighting condition was compared to a condition
with highlights from a simulated CAD system with TP and FP rates of 0.75 and
0.10, respectively. When candidate targets were masked by the noise, highlights
improved sensitivity at the highlighted locations but decreased sensitivity in
non-highlighted locations. Fixation tracking confirmed that highlights narrowed
the search area.

The clear message from these studies is that there is no free lunch: hard
highlights have costs that can outweigh benefits \cite{DOrsi2001}.
Trustworthiness of the CAD is key \cite{Dzindoletetal2003}. When the CAD
appears unreliable, individuals expend cognitive resources actively ignoring
the assistance; when the CAD seems reliable, individuals may become overly
reliant and overlook non-highlighted targets \cite{ParasuramanRiley1997}.  One
solution is to completely automate the search task, yet despite the recent
evolution of machine learning methods \cite{Hinton}, medical and remote-sensing
experts are not yet willing or able to hand the reins to automated methods. The
best we can aim for at present is human-machine synergies.

Soft highlighting has the potential to overcome two key limitations of hard 
highlighting. First, individuals can come to trust a well calibrated but
imperfect classifier because soft highlights indicate graded confidence 
not a binary decision.  Second, individuals can better integrate bottom-up 
guidance from the highlights with their own top-down strategic search
because soft highlights provide graded saliency cues.
% MIKE REMOVED PARAGRAPH BREAK FOR SPACE
The potential of soft highlighting depends on individuals being able to
interpret and leverage confidence measures obtained from a classifier. 
Evidence supporting this ability is found in studies of interactive
CAD systems that can be queried for a suspiciousness or malignancy score
\cite{184,185}. The availability of this information improves discrimination
performance over a traditional CAD system \cite{Hupse2013}.

%INDICATION THAT MORE SUBTLE METHODS WILL HELP:  POINTING TO EXPERT BEHAVIOR 
%A particularly interesting example of hard agent-based highlighting in medical
%imaging is found in Litchfield 2010 [123]. It has been demonstrated, for
%example in [110], that radiologists will fixate for longer periods of time even
%on lesions in medical images that are missed. This caused Litchfield in [123]
%to explore the question of what happens when a radiologist, in this case a
%novice, scans chest x-rays looking for pulmonary nodules by following the path
%taken by an expert radiologist. The result is that the gaze path of the expert,
%in this case the agent, when shown to the novice increases the novice’s ability
%to locate lesions. Figure 1.25 illustrates such a path and how it is
%superimposed over the chest x-ray image.  This form of hard agent-based
%highlighting was demonstrated to improve the performance of viewers searching
%for lung nodules.

%"HARD" highlighting and uncertainty
%Some CAD systems, especially those used in research, are interactive and allow users to query locations in order to be told what the likelihood is that a lesion is in that region [186]. However, even with these systems the highlight is hard and consists of an outline, perhaps generated by region growing, indicating a likely location for further examination.

\section{Experiments with soft highlighting in synthetic displays}

We report on a series of seven experiments investigating highlighting in visual
search. The experiments are aimed at determining whether individuals exploit
the information in soft highlights, and whether this information leads to
superior performance over the use of hard highlights and a no-highlight
control.  Experiments 1-4 test synthetic displays and simulated classifiers of
various strengths. Experiments 5-7 test naturalistic images and an actual
state-of-the-art neural network classifier.

\begin{figure}[tbp]
\centering
\begin{tabular}{clclclcl}
(a) &
\includegraphics[width=1.2in,align=t]{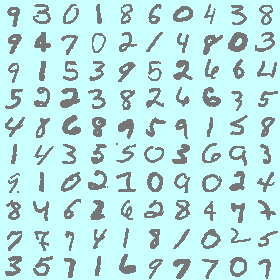} &
(b) &
\includegraphics[width=1.2in,align=t]{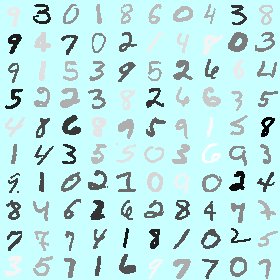} &
(c) &
\includegraphics[width=1.2in,align=t]{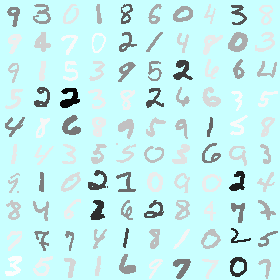} &
(d) &
\includegraphics[width=1.2in,align=t]{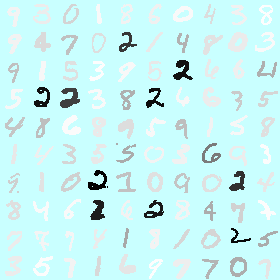} \\
\end{tabular}
\caption{Examples of synthetic digit arrays used in Experiments 1-4. (a)
a control image with no highlighting of targets. (b) highlighting 
with discriminability $\dprime = 0.75$. (c) $\dprime = 3.10$.
(d) $\dprime=5.80$.
\label{fig:synthetic_displays}}
\end{figure}

\subsection{Experiment 1: Search for multiple targets}
In Experiment 1, participants searched a $10\times10$ array of handprinted digits
for all instances of a target---the digit 2.
Figure~\ref{fig:synthetic_displays}a shows a sample array, which consists of
ten tokens of each of the digits 0-9 placed in random locations. Participants
click on each target, causing the target to disappear, and a trial continues
until all ten targets have been localized or until 15 seconds has passed.  We
compare a control condition with no highlighting to conditions in which
elements of the array are highlighted based on simulated classifiers varying in
quality (Figures~\ref{fig:synthetic_displays}b-d).  Soft highlighting is
achieved by modulating the grey level of pixels of an element.  With the light
blue background, dark elements are salient and light elements are difficult
to detect against the background.

We refer to our simulated classifier as a \textit{stochastic oracle-based
classifier} or \textit{SOC}. The SOC generates values in $[0,1]$ that represent
the probability or confidence level that a given element is a target.  The term
`oracle' refers to the fact that we know the true class label of each element
chosen for our displays. The term `stochastic' refers to the fact that an 
element's confidence level is drawn from a probability distribution---either
$\mathrm{Beta}(\theta,1)$ if the element is a target, or the complementary
distribution, $\mathrm{Beta}(1,\theta)$ if the element is a nontarget.
With $\theta=1$, both distributions are uniform,
and the level provides no information about the true class label. As $\theta
\to \infty$, an element's confidence approaches a deterministic level of 1.0
for targets and 0.0 for nontargets.  The quality of the SOC is thus specified
by $\theta$.  Because $\theta$ is not an intuitive measure, we use two other
measures to describe the quality: the \textit{equal error rate}, the error rate
based on a decision threshold that matches TP and TN rates, and $\dprime$, a
measure of discriminability based on signal-detection theory \cite{Green1966}.
Table~\ref{tbl:classifier_quality} shows the three equivalent measures of
quality of SOCs used in Experiments 1-4. Hereafter, we use $\dprime$ to refer
to the quality, and \soc{\dprime} to refer to an SOC with a given
$\dprime$.  The $\dprime$ value associated with
Figures~\ref{fig:synthetic_displays}b-d is presented in the Figure caption.
Note that even for values of $\dprime$ traditionally considered large---say,
greater than 2.0---not every target pops out and not every nontarget is
concealed. The existence of a threshold that reliably discriminates targets
from nontargets does not imply that targets and only targets will be
perceptually salient.

\begin{table}[tb]
\tbl{Three equivalent measures of the quality of stochastic oracle-based classifiers (SOCs) used in Experiments 1-4.}{
\begin{tabular}{|c||c|c|c|c|c|c|c|c|}
\hline 
$\theta$ & 1.00 & 1.50 & 1.74 & 1.90 & 2.33 & 4.00 & 9.00 & 19.00 \\
\hline 
EER & 0.50 & 0.36 & 0.30 & 0.27 & 0.20 & 0.06 & 0.002 & 0.000 \\
\hline 
$\dprime$ & 0.00 & 0.75 & 1.00 & 1.25 & 1.69 & 3.07 & 5.77 & 9.24  \\
\hline 
\end{tabular}}
\label{tbl:classifier_quality}
\end{table}

In Experiment 1, our goal was to determine the classifier quality necessary
to obtain a benefit of soft highlighting over no highlighting. If individuals
are sensitive to graded highlights, then even relatively poor classifiers
may boost performance. 

\subsubsection{Stimuli and design}
Displays were generated for six experimental conditions in which soft
highlighting is based on \soc{\dprime} for $\dprime \in
\{0.00, 0.75, 1.69, 3.07, 5.77, 9.24\}$ and for a control condition in which
all elements are gray.  Each display is a $10\times10$ array of handprinted  
digits consisting of 10 instances of each of the digits 0-9, randomly selected
without replacement from the MNIST data set \cite{Lecun1998}.
Each digit consists of $28\times28$ pixels whose values are thresholded 
%%% thresholded at intensity == 30
such that each pixel is either off or on. The background color has RGB values
on a 0--1 scale of $(.79, 1.0, 1.0)$.  The foreground color of a digit with
assigned classifier confidence $\rho \in [ 0,1 ]$ is set to an RGB value
converted from Lab space with lightness $1-\rho$ and $a=b=0$.  Conversion from
Lab to RGB for display passes through the XYZ color space.  This color space is
based on a reference illuminant of which there are several possible choices.
For our experiments, we used the D55 reference illuminant.  The use of the
Lab-to-RGB lightness conversion was intended to match an element's perceptual
salience to the classifier confidence.
% MIKE REMOVED PARAGRAPH FOR SPACE
The experiment consisted of 42 trials in six blocks, with one trial from
each of the 7 conditions in each block. Trials within a block were randomized.
There were no breaks between blocks.

\subsubsection{Participants}
In all experiments, participants were recruited via Amazon's Mechanical Turk 
platform for pay, at a rate of \$8/hour. Thirty
five adults with normal color vision were accepted into the study and 
25 completed all trials. The remainder either quit the experiment 
voluntarily or were rejected for failing to satisfy a requirement that 
they maintain focus on the browser window through the course of the experiment.
We did not record the manner of termination.

\subsubsection{Procedure}
%All data and images for the experiment were downloaded to a participant's
%computer before the start of the experiment. 
Each trial
began by presenting the $10\times10$ array centered on the participant's monitor.
%so that each trial appeared in the same location on the screen. % 
Participants were
instructed to click on all targets (the digit 2) in the display. When a target
was clicked, it was removed from the display and replaced with the background
color. Clicking on a nontarget had no effect. If participants failed to 
find all 10 targets within 15 seconds, 
the trial terminated and they were informed how many targets they
had found.  At the end of each trial, a `next' button appeared below the array
and the experiment paused until the participant clicked to proceed. The
requirement to click on the button ensured that each trial began with the mouse
in a known position relative to the array. If participants failed to find at
least two targets or clicked on more than six nontargets, they were rejected and
the experiment terminated. These restrictions ensured that participants
maintained focus on the experiment window until completion of the experiment.

\subsubsection{Results}

\begin{figure}
\centering
\setlength\tabcolsep{1pt}
\begin{tabular}{clcl}
(a) &
\includegraphics[width=3in,align=t]{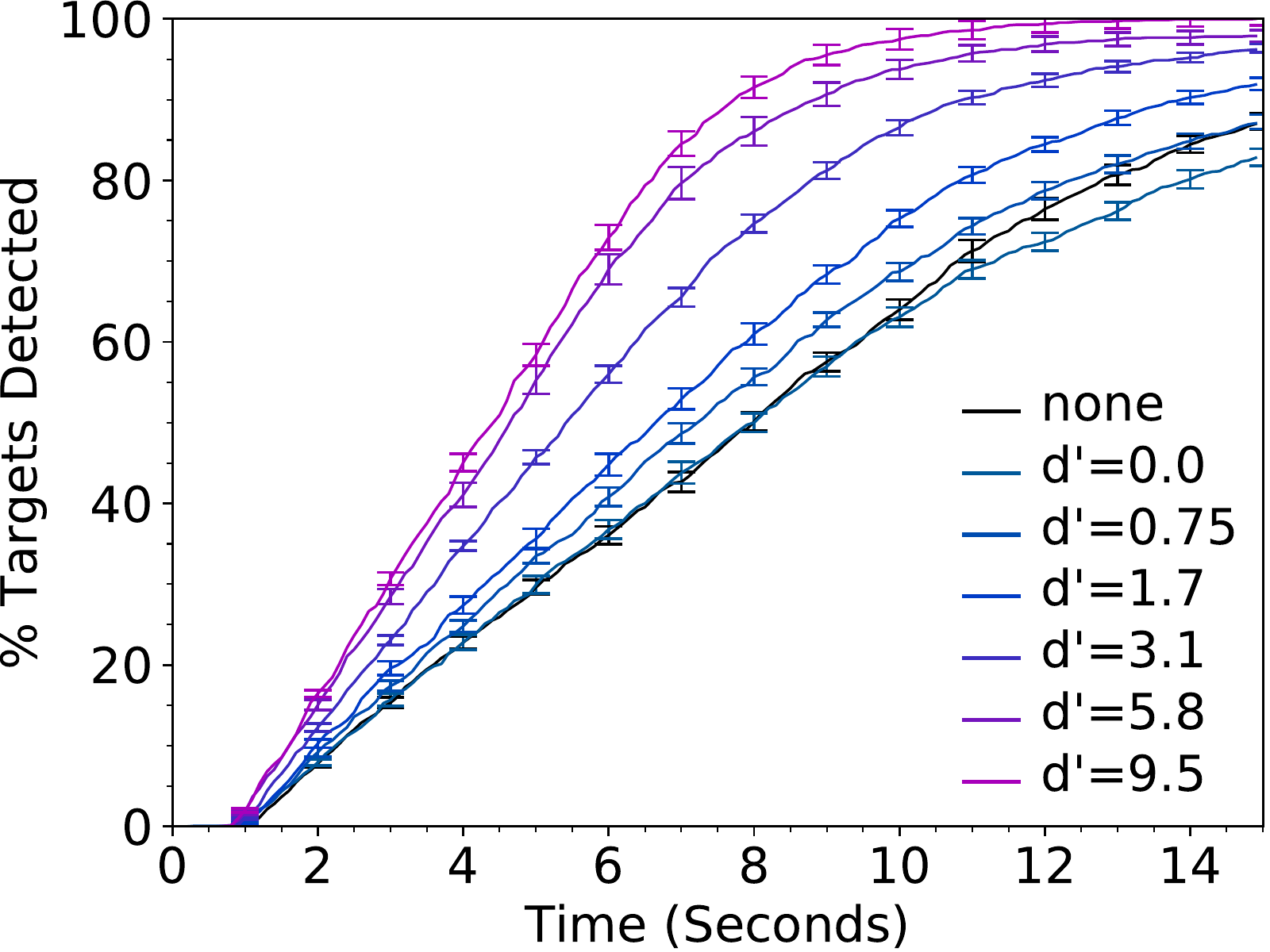} &
(b) &
\includegraphics[width=3in,align=t]{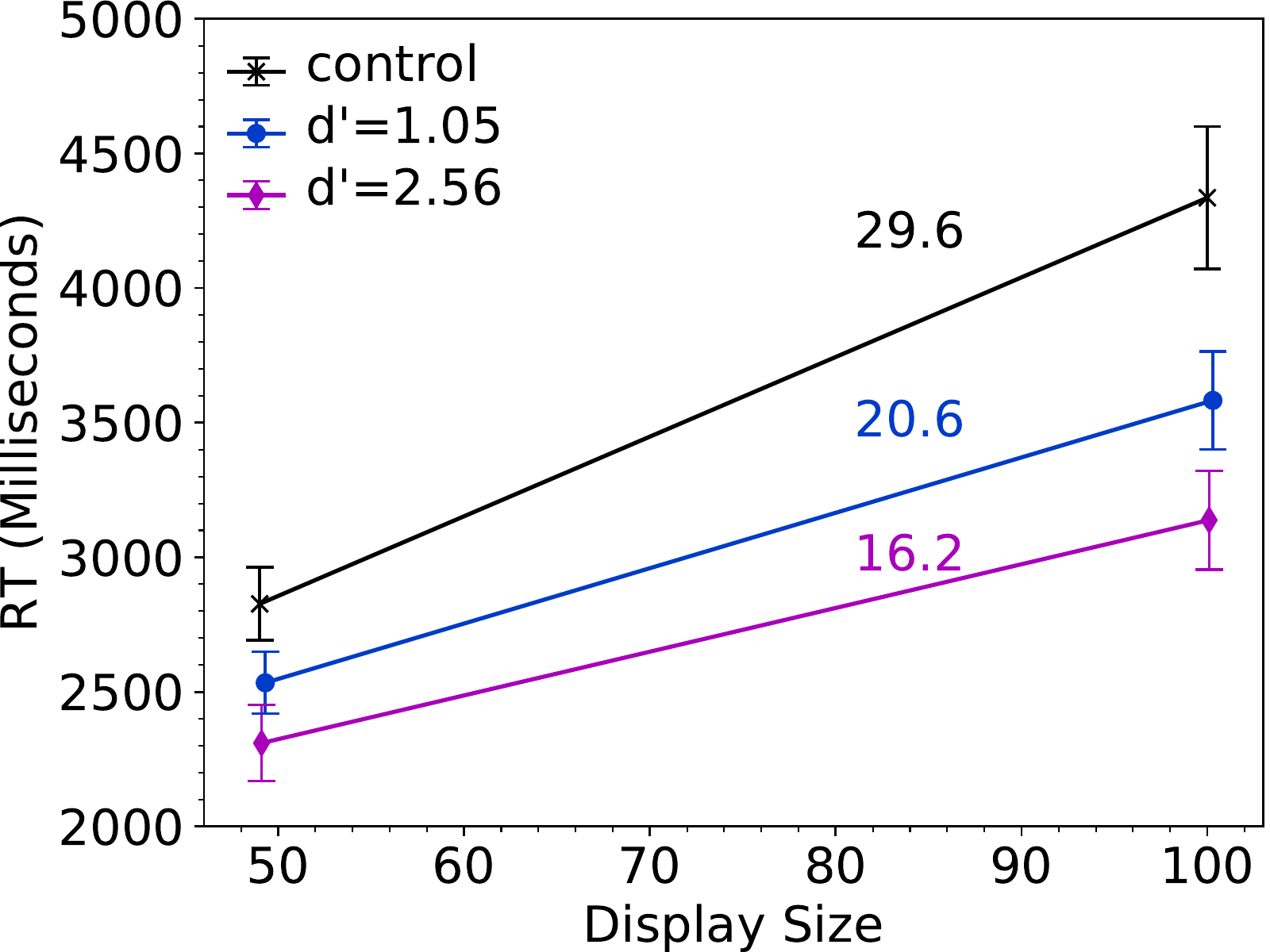} \\
\end{tabular}
\caption{(a) Experiment 1: Percentage of the ten targets 
detected as a function of time and condition. Error bars
indicate $\pm1$ standard error of the mean corrected for between-participant
variability according to the procedure in \cite{Masson2003}.
(b) Experiment 2: Mean target-detection latency for the two display sizes
and three highlighting conditions
}
\label{fig:E12_results}
\end{figure}

For each participant and each condition, we compute the percentage of the ten
targets detected as a function of time from the onset of the
stimulus array. Figure~\ref{fig:E12_results}a shows the mean across
participants for the no-highlighting control condition (black) and the six
experimental conditions (shaded from blue to purple in increasing order of
classifier quality).  A curve that rises quickly and then asymptotes at 100\%
is indicative of an easy search; a curve that rises slowly and doesn't reach
100\% by the end of the 15-second trial is indicative of a difficult search.
Error bars in the Figure are corrected for between-participant variability
according to the procedure in \citeN{Masson2003}.
Nontarget clicks are infrequent, with a mean rate of 0.19 per trial across
participants and conditions.
%$0.19\pm 0.03$

Participants are sensitive to classifier quality: the soft
highlighting curves do not cross over one another, indicating that the
superiority of one classifier over another is consistent across time.  
All soft highlighting conditions with \soc{\dprime > 0} lead to more
efficient search than the no-highlighting condition.  We are surprised by the
finding that even \soc{0.75} is helpful; our informal
polling of colleagues led to the prediction that only the highest quality
classifiers would facilitate search.  Participants are able to exploit the
signal provided by a rather weak classifier.  The pattern of results appears to
rule out two alternative hypotheses: first, that relatively weak classifiers
draw attention to the wrong elements and therefore suffer relative to the
control; and second, that soft highlighting leads to an early advantage---due
to the appropriately highlighted elements---but a late cost---due to the
inappropriately highlighted elements---relative to the control.

To conduct a more quantitative analysis, we computed the median time for each
each participant to find \textit{five} targets in each condition and performed
paired $t$ tests between conditions.  Control, \soc{0}, and \soc{0.75} had mean
times of 7.54, 7.67, and 7.38 sec, respectively, though none of these pairwise
differences were reliably distinguishable. Using median time to find
\textit{eight} targets, \soc{0.75} is faster than control (10.67 versus 11.41
sec, $t(24)=3.36, p=.0025$) and faster than \soc{0} (10.67 versus 11.52 sec,
$t(24)=2.56, p=.017$).  Focusing instead on the number of targets detected
halfway through the trial, at 7.5 sec, \soc{0.75} obtains a greater proportion
of detections than control (0.54 versus 0.49, $t(24)=3.41, p=.002$) but not
\soc{0} (0.54 versus 0.51, $t(24)=1.40, p=.17$). We have given
ourselves license to fish through the data, and one should therefore
interpret the findings cautiously, there does seem to be a preponderance
of evidence that \soc{0.75} highlights are sufficient to facilitate search. 

%Control and SOC0   :        4.8800000       5.1200000      -1.4446302      0.16149190
%Control and SOC0.75:        4.8800000       5.4400000      -3.4122667    0.0022875432
%SOC0 and SOC0.75   :        5.1200000       5.4400000      -1.3979265      0.17492275

Because the experimental conditions are intermixed, participants cannot
anticipate the trustworthiness of the classifier on any trial. However, due to
our simulation of classifier outputs, an easily discernible statistic---the
spread or variance of the confidence levels---is a reasonable proxy for
classifier quality: the more uniform the spread, the less reliable the
classifier.  Thus, we do not expect results would change much with a blocked
design.

%The consistency of the ordering of conditions over time facilitates a clean
%interpretation of our results.

%
% TODO
%
% RON, ideally you'd have an analysis like that we've discussed in the
% past -- a more classic analysis in which you obtain a latency for each
% condition and then do t-tests to see whether the latencies differ.
% For example, you could compute for each trial the mean or median time to find
% 5 targets for each subject in each condition, and then do t-tests comparing
% control to d'=0, d'=0 to d'=0.75, etc.  you should see reliable differences
% between every pair of conditions except for the control vs. d'=0.
% This quantitative analysis, backed by stats, would support the qualitative
% analysis based on the curves in the figure. I won't be able to get this
% work published without quantitative analyses.

\subsection{Experiment 2: Search for a single target in variable-sized displays}

Traditional studies of visual search utilize
displays with zero or one targets and a 
variable number of distractors [e.g., \citeNP{TreismanGelade1980}].
Experiment 2 follows this paradigm with
single-target displays of two sizes, $7\times7$ or 
$10\times10$.
This design allows for the measurement of a search {\it rate}---the increase in
response time per additional display element.  The speed up from soft
highlighting observed in Experiment 1 could be due either to an increased
search rate or to factors that are unrelated to display size, such as early
perceptual processing, initial segmentation of the display, or motor
preparation.  These two possibilities can be distinguished via an affine model
relating response latency, $t$, to display size, $s$: $t = \alpha + \beta
s$, where $\beta$ is the search rate, and $\alpha$ is display-size
independent processing.

\subsubsection{Stimuli and design}
Each stimulus array consists of a single target---the digit 2---placed in 
a random cell, with the remaining cells filled at random with nontarget 
digits. No digit token appeared more than once in the course of the experiment.
The $7\times7$ arrays occupied the same area as a $7\times7$ portion of the
$10\times10$ arrays and were centered on the screen. 
% MIKE REMOVED PARAGRAPH BREAK FOR SPACE
The experiment consisted of 15 blocks of 6 trials, for a total of 90 trials.
Each block had exactly one trial that crossed display size---$7\times7$ or
$10\times10$---with highlighting condition---control, \soc{1.05}, and \soc{2.56}.
As in Experiment 1, order of trials within a block was randomized, and no
break or indication was given at the end of a block.

\subsubsection{Participants}
Of 56 participants who enrolled, 37 completed all trials;
the remainder were either rejected or terminated their participation 
voluntarily. All had normal color vision.

\subsubsection{Procedure}
Participants were instructed to locate and click on the target. Clicks on
nontargets were ignored. To ensure adherence to instructions, participants were
rejected if they failed to find the target within 45 seconds or if they clicked
on more than six nontargets in a trial.  At
the end of each trial, response latency was displayed and a 'next' button lit
up that would initiate the following trial when the participant was ready.

\subsection{Results}
Figure~\ref{fig:E12_results}b shows the mean latency for target detection for
each of the three highlighting conditions---control, \soc{1.05}, and
\soc{2.56}---and the two display sizes---$7\times7$ and $10\times10$. An ANOVA
indicates a main effect of highlighting condition ($F(2,72)=14.4, p< 0.001$)
and display size ($F(1,36)=127.6, p < 0.001$). Critically, the interaction
between highlighting and display size is reliable ($F(2,72) = 5.18, p= 0.008$).
The search slope in the control condition appears steeper than in the \soc{1.05}
condition, which in turn is steeper than in the \soc{2.56} condition. Thus,
highlighting reduces the time to search for each item in the display,
consistent with the notion that highlighting guides attention to relevant
locations.

The slope of the best fit line to each condition is shown in 
Figure~\ref{fig:E12_results}b. Performing pairwise comparisons, both
the nonparametric two-sided Wilcoxon signed-rank test and paired t-test
show a marginal difference between the slopes for control versus \soc{1.05}
($w(36)=1550, p=.079$; $t(36)=1.89, p=.067$) and a reliable difference for
control versus \soc{2.56} ($w(36)=1652, p=.004$; $t(36)=3.18, p=.003$). No
difference was observed between \soc{1.05} and \soc{2.56} ($w(36)=1492, p=.24$;
$t(36)=1.18, p=.26$). Our findings are thus consistent with the claim that
highlighting increases the efficiency of search.

% RON: haven't seen the stat test for intercepts
%there is a statistically significant difference in slopes between the 
% highlighted and control conditions.  This also applies to the intercepts.

% RON: from the curve fits, you should be able to determine the intercept
% values, and you can ask if the intercepts are any different.  That is, for
% each subject compute that subject's intercept for each of the 3 conditions.
% Then do a 3-way anova to ask if there are differences among the three 3
% intercept values. 
% Even though the slope differences are most interesting, intercept differences
% must be reported and have some meaning or at least need some interpretation.

\subsection{Experiment 3: Comparing soft and hard highlighting}

\begin{figure}
\centering
\includegraphics[width=5in]{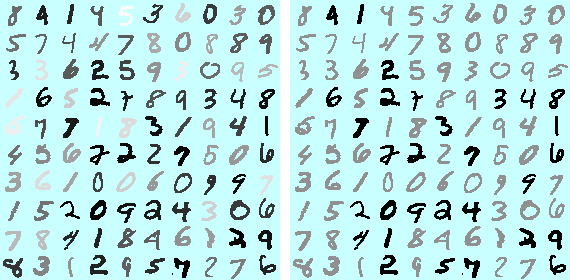}
\caption{Experiment 3, sample stimulus arrays: (left) soft highlighting
via \soc{1.05}; (right) hard highlighting obtained by thresholding \soc{1.05}
confidence levels.}
\label{fig:expt3_stim}
\end{figure}

Having established that individuals are able to exploit the signal
in soft highlights to guide attention, we turn to the key issue of how
soft highlighting fares relative to hard highlighting. Three classifiers
were studied: \soc{0.75}, \soc{1.00}, and \soc{1.25}.  We selected this set of 
relatively weak classifiers to further investigate the relationship between 
classifier quality and human performance.  Soft highlighting
was based on the SOC output, as in earlier experiments. Hard highlighting
was based on thresholding the SOC output at 0.5, yielding displays such
as that shown in the right panel of Figure~\ref{fig:expt3_stim}.

\subsubsection{Stimuli and design}
Experiment 3 used digit arrays identical to those used in Experiment 1, with
ten targets per array.  In addition to a control condition with no
highlighting, 6 classifier conditions were formed by the Cartesian product of
highlighting type---soft, hard---and classifier quality---\soc{0.75}, \soc{1.00},
and \soc{1.25}.  The experiment consisted of six blocks, each with one instance
of the 7 conditions, for a total of 42 trials. For hard highlighting,
classifier confidence is thresholded at 0.5. Elements above and below this
threshold are set to a display intensity equivalent to a confidence of 1.0 and
0.2, respectively. The level of 0.2 was selected to ensure that the digits were
readily perceptible but were less salient than the highlighted digits. We
conducted pilot studies in which we performed hard highlighting by circling
elements or changing the background color of the highlighted element, but
decided to match the appearance of displays with soft highlighting.

\subsubsection{Participants}
Of 54 participants who enrolled in the experiment, 41 completed all trials.
The remaining 13 either quit the experiment on their own or were rejected. We
did not record the reason for their failure to complete.

\subsubsection{Procedure}
Experiment 3 was run in a manner identical to Experiment 1 with identical
instructions.

\subsubsection{Results}

\begin{figure}
\centering
\setlength\tabcolsep{2pt}
\begin{tabular}{clcl}
(a) &
\includegraphics[width=3in,align=t]{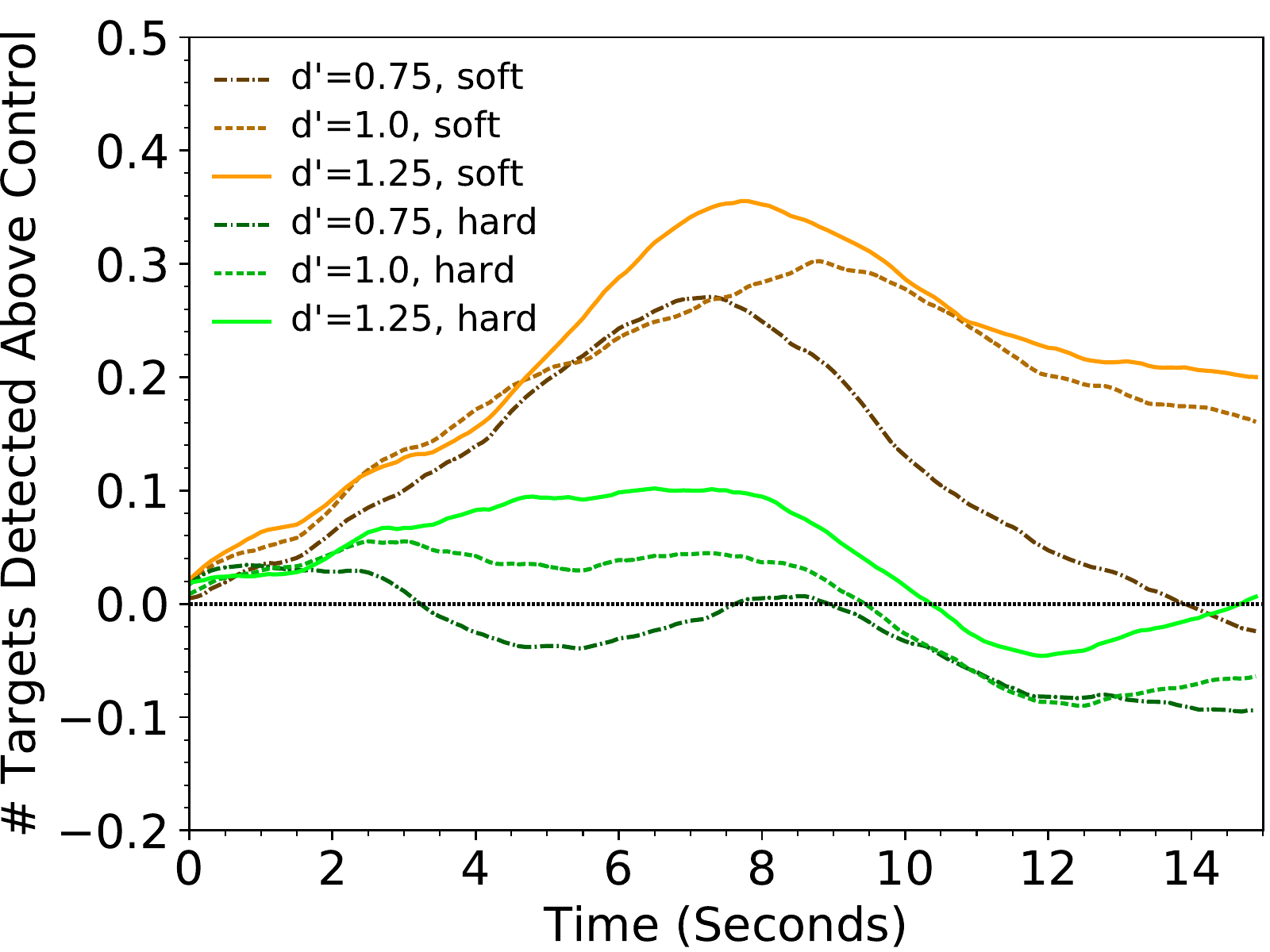} &
(b) &
\includegraphics[width=3in,align=t]{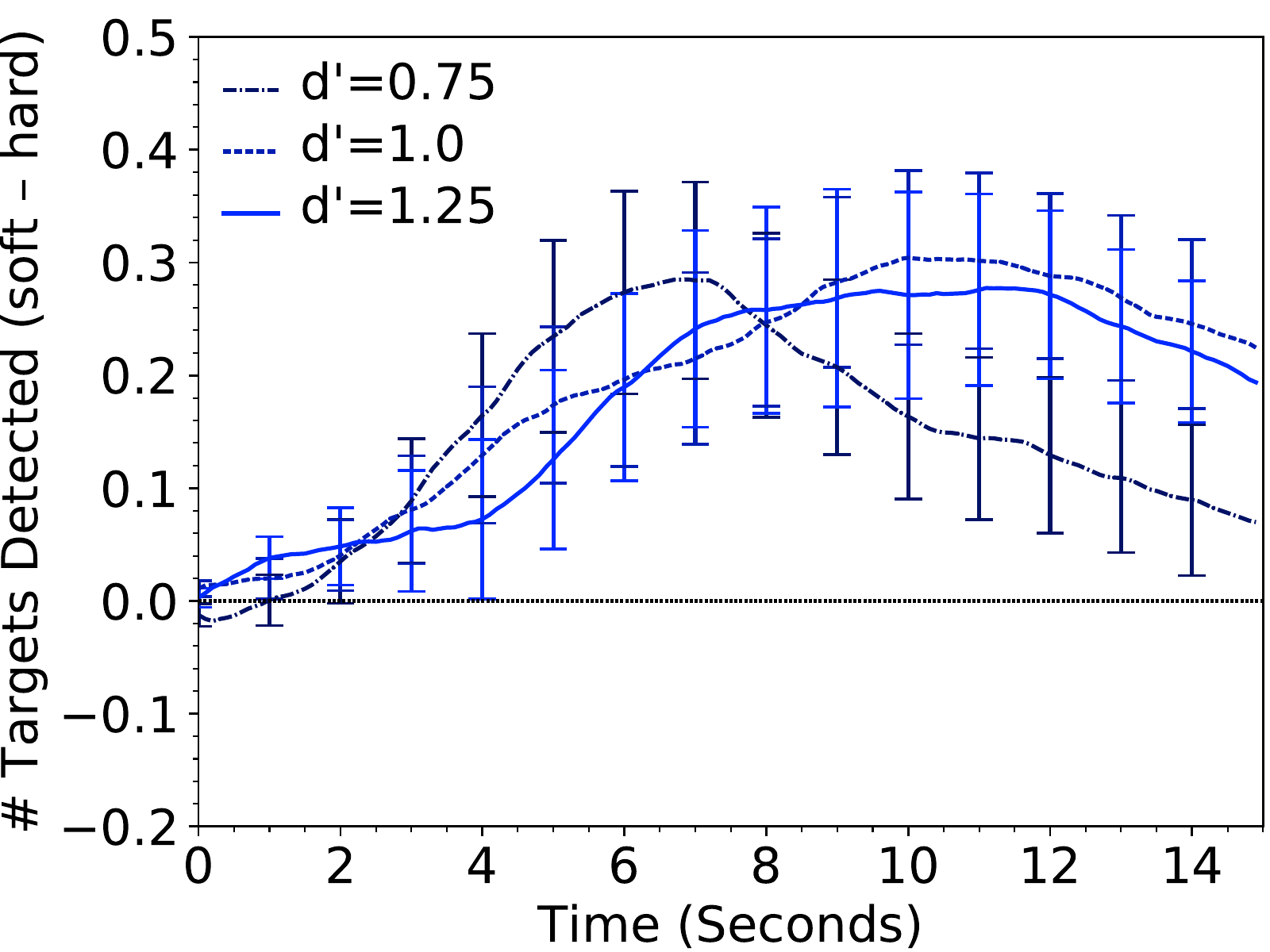} \\
\end{tabular}
\caption{Experiment 3. (a) The mean number of targets located across all
participants for each condition minus the mean number of targets located across all
participants for the control condition of no highlighting as a function of time.
Values above zero mean that more targets were detected relative to the control
condition while values less than zero mean fewer targets were detected. 
(b) The difference between the mean number of targets detected by that time for the soft and hard highlighting conditions for corresponding SOCs. Positive
values reflect better performance with soft than hard highlighting. Error bars
are calculated using a between-participant variability correction \cite{Masson2003}.}
\label{fig:E3_results}
\end{figure}

For each participant and condition, we compute the number of targets detected
as a function of time within the 15-second trial. The curves have the same
general shape as the data from Experiment 1 (Figure~\ref{fig:E12_results}a),
but to accentuate differences between conditions, Figure~\ref{fig:E3_results}a
plots the difference in the number of targets detected between each 
highlighting condition and the control condition.
The control is depicted as a black dashed line at
the baseline of zero. Values above zero indicate that more targets
were located by that instant than in the control condition. 
Qualitatively, hard highlighting appears to have a small
benefit relative to the control for \soc{1.25} between 3 and 9 sec, though
the benefit for \soc{1.00} and \soc{0.75} appears minimal.
In contrast, soft highlighting appears to
have a larger advantage relative to the control for all three classifier
strengths and for most of the duration of the trial.
% Note that a 3-second wide smoothing window is applied.

Figure~\ref{fig:E3_results}b shows the difference between the mean number of
detected targets in corresponding soft versus hard highlighting conditions,
as a function of time. Values of the curve greater than zero indicate 
that at a particular instant of time, participants found more targets
with soft highlighting than with hard. An advantage is observed for
soft highlighting for all three SOCs.

To perform a quantitative analysis, we divided the first 12 seconds of each
trial into 4 bins of 3 seconds. We computed the number of targets found within
each bin and performed a three way ANOVA with participant as the random factor and
classifier quality, highlighting type, and time window (0-3, 3-6, 6-9, and 9-12
seconds) as three within-participant independent variables.  We observe a main
effect of highlighting condition ($F(1,40)=10.213$, $p=0.003$), with soft
highlighting superior to hard highlighting. We also observe a main effect of
time ($F(3,120)=933$, $p<0.001$), which simply reflects the fact that more
targets are found around 6 or 9 sec than around 3 or 12 sec. No main effect of
classifier quality is observed ($F(2,80)<1$), not terribly surprising
given the small range of $d'$ tested. No interactions involving these three
factors are significant at the 0.05 level.

The effect size of soft versus hard highlighting was assessed at 6 and 9
seconds for each of the 3 classifier qualities. The effect sizes range
from small to medium: Cohen's $d$ at 6 sec is 0.53, 0.24, and 0.25 for 
\soc{0.75}, \soc{1.00}, and \soc{1.25}, respectively; 
and at 9 sec 0.22, 0.32, and 0.27.
%6 sec & $d'=0.75$ & Cohen's $d= 0.528$ \\
%6 sec & $d'=1.00$ & Cohen's $d= 0.243$ \\
%6 sec & $d'=1.25$ & Cohen's $d= 0.248$ \\
%9 sec & $d'=0.75$ & Cohen's $d= 0.219$ \\
%9 sec & $d'=1.00$ & Cohen's $d= 0.324$ \\
%9 sec & $d'=1.25$ & Cohen's $d= 0.265$ \\
For a given SOC, our results indicate that soft 
highlighting, which leverages the classifier's graded output, supports
human visual search better than hard highlighting, which thresholds the
classifier output.

\subsection{Experiment 4: Search for a variable number of targets}

Experiments 1-3 show that even a weak classifier can boost human performance 
when soft highlighting is used.  However, because targets are
present on every trial and the number of targets is known, participants
continue searching until all targets are found or time is exhausted.
Consequently, we have no evidence concerning the effect of highlighting on an
individual's decision to quit searching, and therefore, on the possibility of
missed targets. As discussed in the introduction, the medical literature 
suggests that hard highlighting can increase the rate of target misses.

To explore the effect of soft highlighting on target misses, participants
in Experiment 4 searched $10\times10$ arrays that contained
between zero and two targets. They were instructed to click on each target and
to press a `done' button to terminate a trial when they were
confident that no targets remained.  Nontarget clicks were ignored.  

\subsubsection{Stimuli and design}
Each display contained 0, 1, or 2 targets, the digit `2'. The remainder of the
array was filled with randomly chosen digits.  Five highlighting conditions
were included in the design: a control with no highlighting, \soc{0},
\soc{1.69}, \soc{3.07}, and \soc{5.77}.  The experiment consisted of 80 trials,
the first five of which were practice and were not used in analyses. The
practice trials included one with no targets present and 2 each with 1 and 2
targets present and appeared in random order.  The 75 subsequent trials were
presented in 5 blocks of 15. The trials within a block were formed through the
Cartesian product of five highlighting types and 0-2 targets.  The sequence of
trials was such that across the experiment there was exactly one trial per
highlighting type $C$ with $T$ targets following a previous trial of type $P$.
This led to 5 levels of the current type times 5 levels of the previous type
times 3 levels of number of targets to arrive at 75 trials total.

\subsubsection{Participants}
Of 50 participants who enrolled in the experiment, 31 completed all trials and
19 quit the experiment on their own as no participants were rejected
based on performance. 

\subsubsection{Procedure}
% REMOVED FOR SPACE
%As in Experiments 1-3, once a target was clicked it was removed from the
%display and replaced with the background color. 
Each trial continued until the
participant clicked `done'. There was no upper limit on the time
for a trial.  Because this task was more difficult than in the previous
experiments, we provided feedback after each trial to ensure participants
remained vigilant and motivated. This feedback consisted of a smaller version of
the digit array with the actual targets colored green if they had been detected 
or red if they had been missed. A buzzer sounded if any target was missed.

\subsubsection{Results}

The analysis of this experiment is more complex than the analyses of the
previous experiments due to the variable number of targets and the possibility
of a speed-accuracy trade off.

\begin{figure}
\centering
\setlength\tabcolsep{3pt}
\begin{tabular}{clclcl}
(a) &
\includegraphics[width=1.9in,align=t]{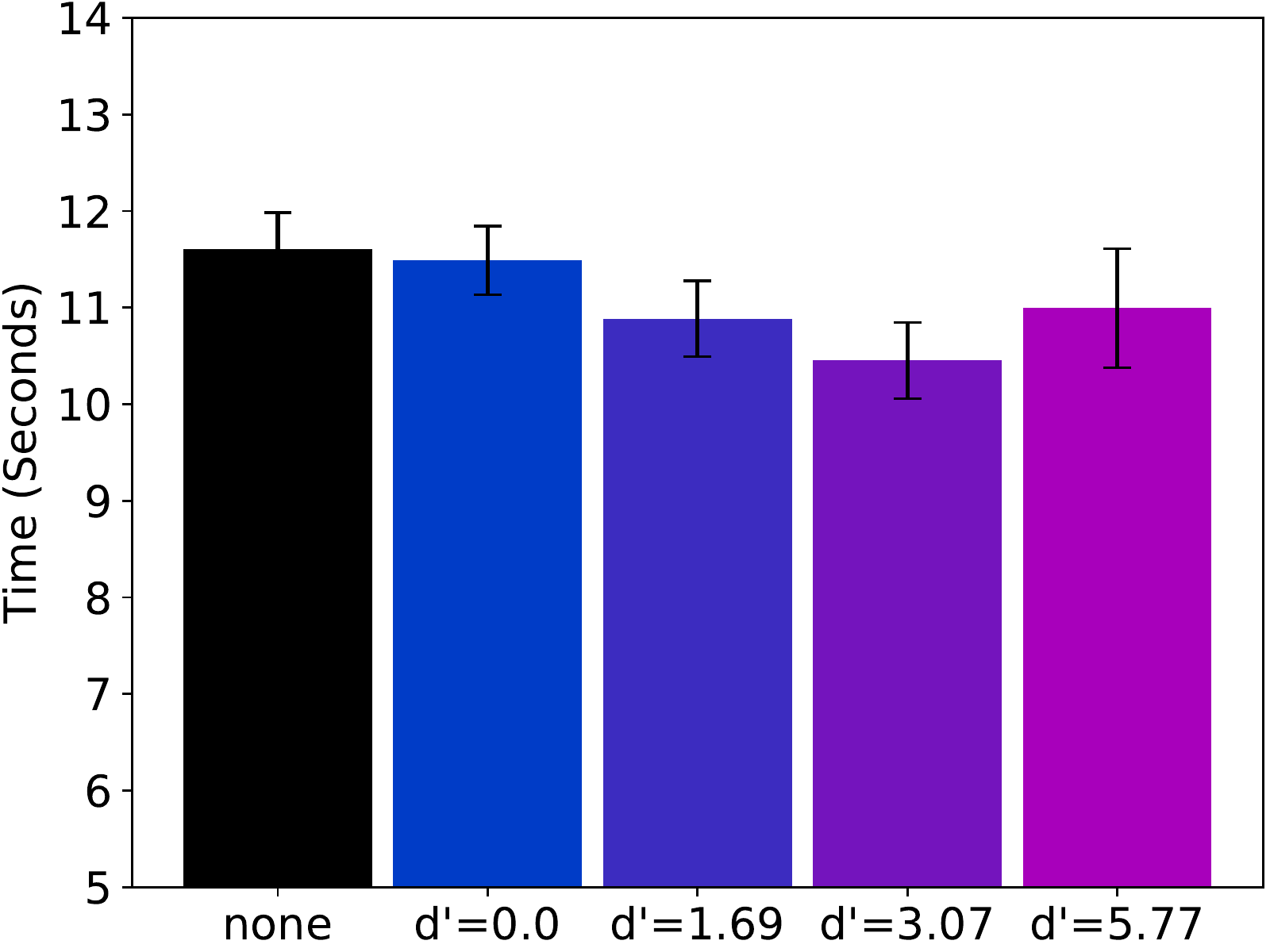} &
(b) &
\includegraphics[width=1.9in,align=t]{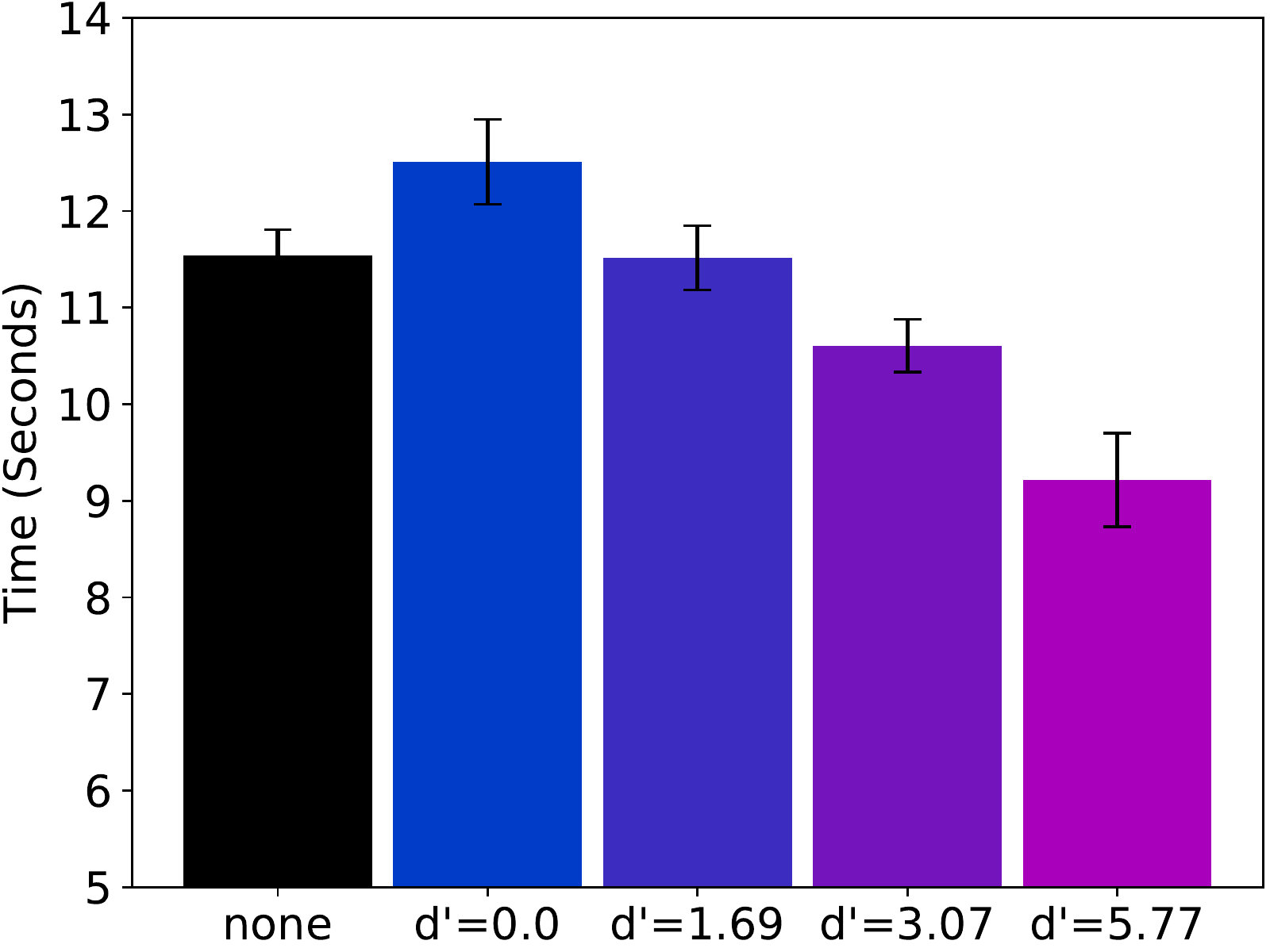} &
(c) &
\includegraphics[width=1.9in,align=t]{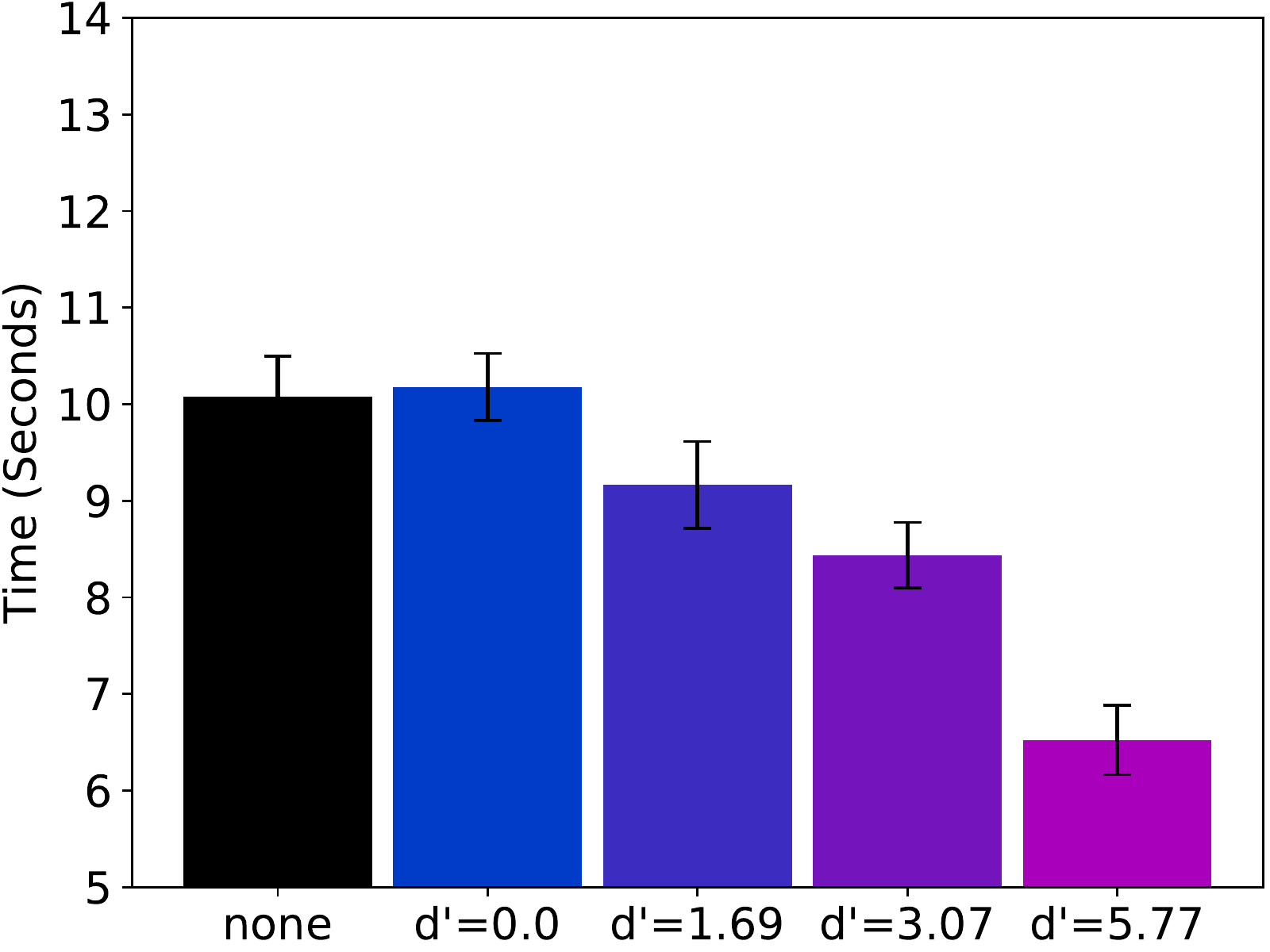} \\
\end{tabular}
\caption{Experiment 4. Mean latency to terminate trial after correctly locating
all targets in (a) zero-target displays, (b) one-target displays, and (c)
two-target displays.  Error bars are calculated using a between-participant
variability correction \cite{Masson2003}.
}
\label{fig:E4_results_speed}
\end{figure}

We first examine zero-target displays and ask how highlighting influences the
time to determine that no target was present and to terminate the trial.  We
compute the median time to terminate for each participant and condition.
Figure~\ref{fig:E4_results_speed}a shows the mean of this statistic across
participants. An ANOVA reveals no main effect of condition ($F(4,120)=0.89$),
indicating that highlighting did not influence the time to
terminate a trial.

\begin{figure}
\centering
\setlength\tabcolsep{3pt}
\begin{tabular}{clcl}
(a) &
\includegraphics[width=2.8in,align=t]{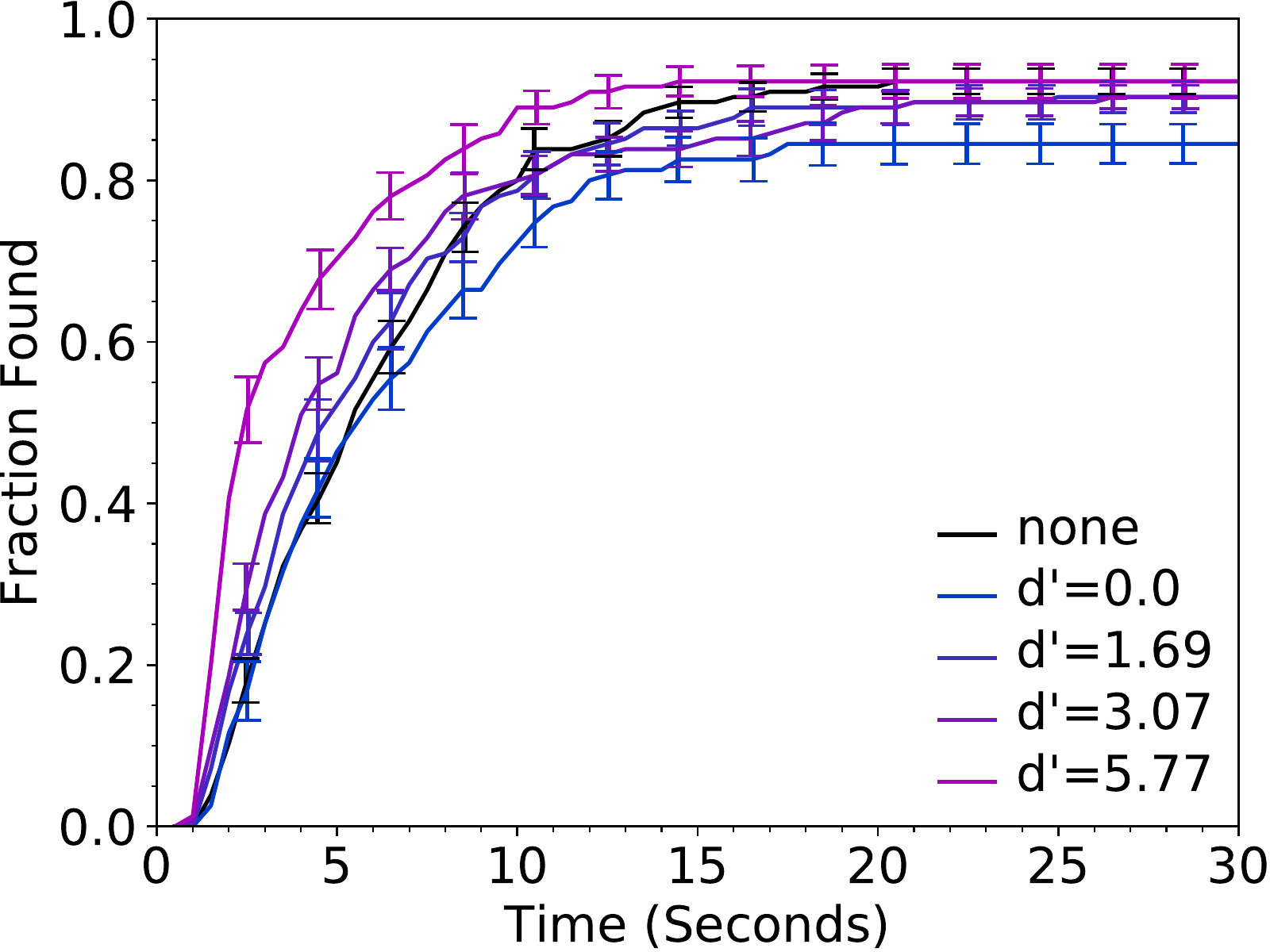} &
(b) &
\includegraphics[width=2.8in,align=t]{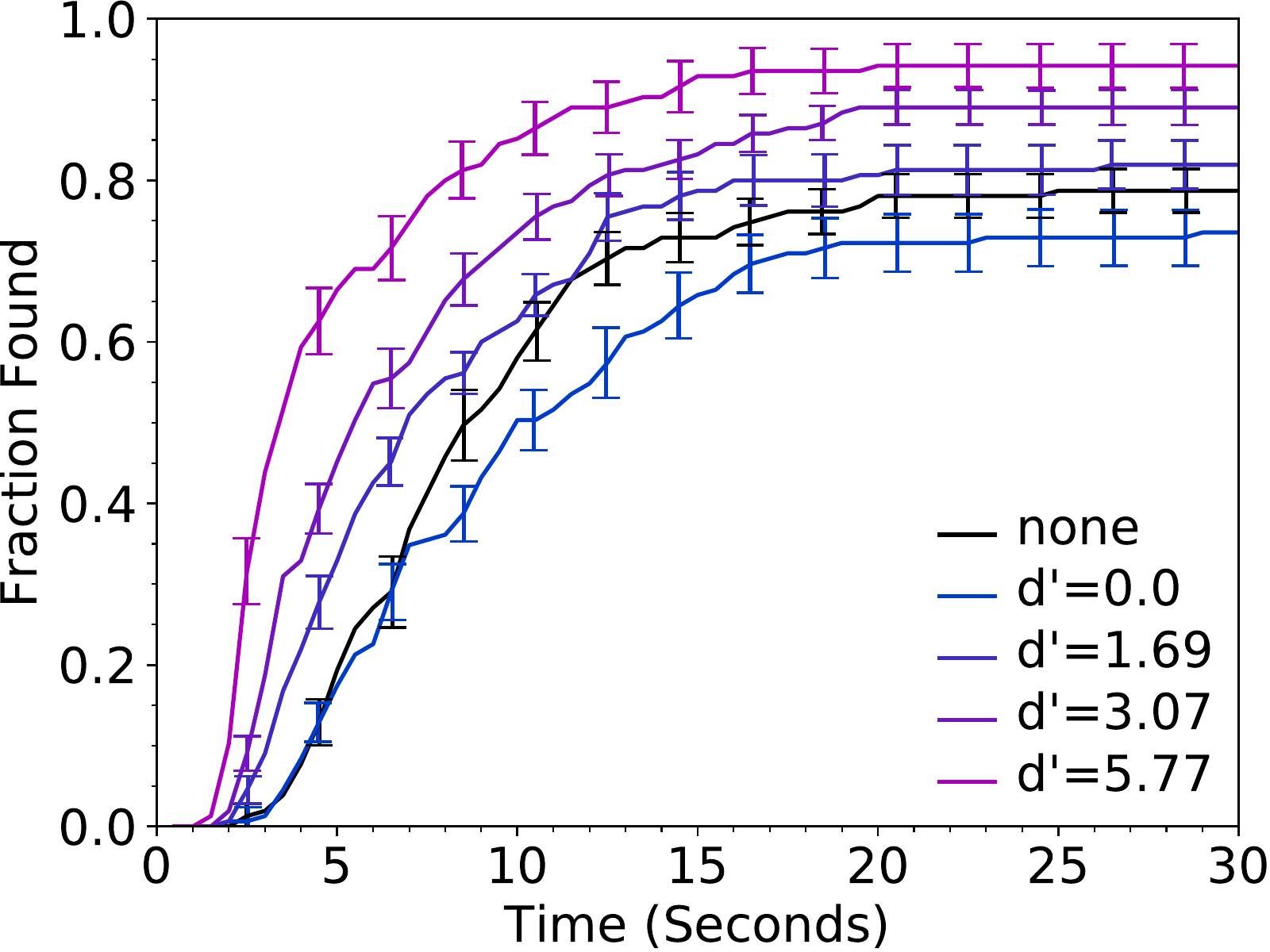} \\
\end{tabular}
\caption{Experiment 4. Mean number of targets detected by highlighting conditions
for (a) one-target displays and (b) two-target displays.  Error bars
are calculated using a between-participant variability correction \cite{Masson2003}.
}
\label{fig:E4_results_acc}
\end{figure}

For one- and two-target displays, detection accuracy asymptotes at 85-93\%
and 75-95\%, respectively (Figure~\ref{fig:E4_results_acc}a,b). 
Differences among conditions do not attain significance for one-target
displays ($F(4,120)=2.10, p=.085$) but are reliable for two-target displays
($F(4,120)=6.18, p<.001$). Both figures show the same trend that higher
classifier qualities yield faster detection and higher asymptotic performance.
Random highlights, as embodied by \soc{0}, appear to impose a cost relative
to no highlighting. 

No hint of a speed-accuracy trade off is observed for target-present displays.
Figures~\ref{fig:E4_results_speed}b,c indicate that time to detect all targets
in one- and two-target displays decreases with increasing classifier quality
(main effect of condition on one target displays: $F(4,120)=8.68, p<.001$; two 
target: $F(4,120)=11.69, p<.001$).  

Across displays with varying numbers of targets, soft highlighting appears to
be an unqualified win.  No speed-accuracy trade offs are induced:
participants miss fewer targets and are faster to complete the task, even with
uncertainty in the presence and number of targets. Experiment 4 involves a more
realistic search task than Experiments 1-3 because of the target uncertainty
and because earlier experiments offered participants no means of
self-terminating a trial without finding all targets.  Experiment 4 strengthens
the results from Experiments 1-3 by showing that, if anything, soft
highlighting leads to earlier termination of search with fewer misses.
However, the conclusions from Experiments 1-4 pertain to tasks in which
target-distractor discrimination is easy.


\subsection{Conclusions from experiments with synthetic displays}

We described four experiments with synthetic displays and simulated classifiers
to explore the influence of soft highlighting on visual search. Although the
search task we studied is simple and artificial, it provides a convenient
vehicle to investigate the influence of classifier quality on performance and
the value of soft versus hard highlighting.  We found that even relatively weak
classifiers boost human performance when the classifier output is
incorporated into search displays with soft highlighting but not with hard
highlighting. We observed across-the-board advantages for soft highlighting
over a no-highlight control. Individuals are sensitive to subtle signals
embedded in noisy classifier output and can leverage these signals to search
more accurately and efficiently.

The displays that we studied, typical of synthetic images in psychological
tasks, are composed of spatially isolated, segmented elements.  Based on
experiments using these displays, one would be reluctant to extrapolate
findings to more complex, naturalistic images and real-world search tasks. In
tasks such as medical and satellite imagery analysis, images are continuous,
display features are not neatly segmented but rather are continuous and 
overlapping and varying in size,
and target-distractor discrimination is challenging even for domain experts.

We therefore conducted a series of experiments involving target search using
naturalistic images: satellite photos of urban and suburban landscapes.
Although participants in our experiments were not experts in analyzing
these images, most individuals have some familiarity with them.
To ensure the task was as
true-to-application as possible, we based highlights on the output of a 
state-of-the-art neural net classifier. In contrast to our SOCs, which generate
outputs that depend only on the target class of a display element, a neural
net's output and errors are systematically related to the image data.  If 
the neural net and humans fail to detect the same targets, then highlighting
is unlikely to boost human performance.

Before turning to the experiments, we first describe the domain, the classifier
we constructed for the domain, and the technique we use for highlighting
satellite imagery.

\section{Analysis of satellite imagery}
\label{sec:classifier}

The analysis of satellite imagery is a critical, time-intensive task performed
by government intelligence, military, forestry, and disaster-relief experts, as
well as biologists and social scientists interested in characterizing human
activity and agriculture. This analysis often requires searching for targets,
e.g., certain types of vegetation, sporting fields, missile silos.

\begin{figure}
\centering
\setlength\tabcolsep{2pt}
\begin{tabular}{clcl}
(a) &
\includegraphics[width=2.75in,align=t]{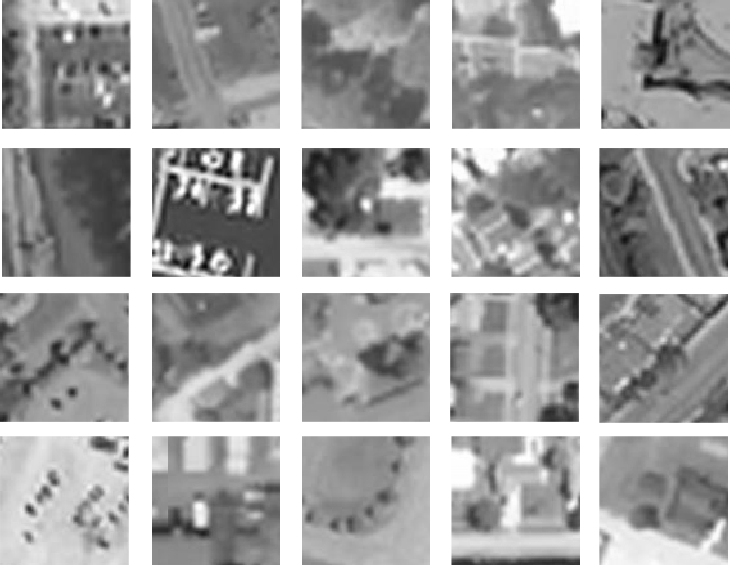} &
(b) &
\includegraphics[width=2.75in,align=t]{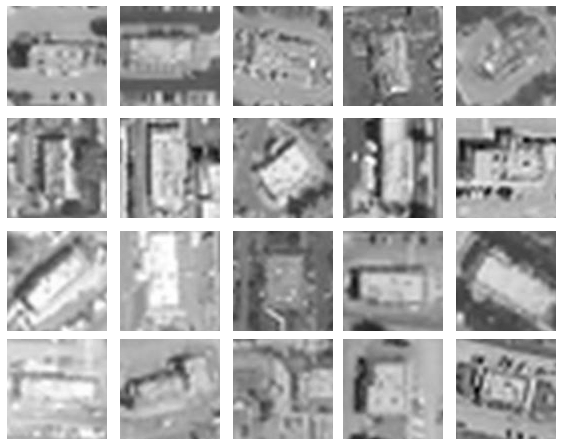} \\
\end{tabular}
\caption{(a) $100\times100$ pixel patches of satellite images (a) not
containing the target, and (b) containing the target}
%{\color{red} MIKE: if we have space
%we could include Fig 5.11 from thesis here.} 
\label{fig:representative_examples}
\end{figure}

We devised a search task that leverages a publicly available point-of-interest
data base containing coordinates of fast-food restaurants operated by a popular
chain (McDonald's) within the United Status.  Using an API from Mapquest, we pulled sets of
satellite images both with and without restaurants at the highest resolution
available.  We converted the pansharpened (color) images to greyscale to follow
the workflow of a typical image analyst who looks almost exclusively at
panchromatic imagery (like a black and white photograph).  We reviewed images
to ensure that a restaurant was present at the specified coordinates and we
rejected restaurants that did not follow the classical design.
Figure~\ref{fig:representative_examples} presents examples of $100\times100$
pixel patches centered on targets and other patches not containing targets.
These patches are large enough to contain the entire restaurant along with
a portion of the surrounding drive-through and parking lot. Variability in the
lay out of the property makes identifying the restaurant challenging.

Stimuli used in our experiments were $1000 \times 1000$ images.
Each contained no more than one target, and the target was randomly offset such
that it could appear anywhere except in the outer 50 pixel wide
border around the image. We obtained 2992 target-present images and
ten times as many target-absent images, the latter being
selected randomly from the 0.15 mile vicinity around the restaurant and
constrained such that no portion of a target was present in the image.  This
proximity-based selection ensures that the target-absent images have roughly
the same density and pattern of trees, roads, and buildings as the
target-present images. We further verified that target-absent images had a
similar character via manual inspection.

\subsection{Constructing a classifier for satellite imagery}

We constructed a state-of-the-art classifier for the restaurant search task.
The classifier takes as input a $100\times100$ pixel patch of an
image and outputs the probability that a target is at least partially contained
in the patch. The classifier is implemented as a convolutional neural net.
Convolutional nets exploit the structure of images via spatiotopically
organized layers of processing units that detect features in local spatial
regions of the layer below. Arranging these feature detectors hierarchically
achieves a transformation from a location-specific, representation of primitive
features (pixels) to a location-independent representation of object
identities \cite{Mozer1991}.  We used Bayesian optimization methods \cite{snoek2014} to select
the specific architecture, which had two convolutional layers and two fully
connected layers. We augmented the the data set by                                  %%% RTK - both were augmented maintaining 1:10 target:nontarget ratio
horizontal and vertical reflections and $90^{\circ}$ rotations.  Cross
validation was used to settle on further aspects of the model and training
procedure. Details of the models, data, and training procedure are described in
\citeN{Kneuselthesis}.

The selected classifier was evaluated on a separate test set which had not been
used during the training process. When thresholded at probability 0.5, the
classifier achieves a TP rate of 0.886 and a TN rate of 0.995. When thresholded
to achieve equal TP and TN rates, the resulting \textit{equal error rate}
(\textit{EER}) is 0.030.  The classifier achieves an \textit{area under the
curve} (\textit{AUC}) of 0.992. (AUC is an alternative measure of
discriminability to $\dprime$ and ranges from 0.5 for a classifier that
performs no better than chance to 1.0 for a classifier that perfectly
discriminates positive and negative examples.  See \citeN{Green1966} for
further details.) Although these statistics seem to suggest that the
classifier is extremely strong, one must take into account that a target can
appear anywhere in a high-resolution image, affording many opportunities
for false positives.

\subsection{Highlighting satellite imagery}

\begin{figure}
\centering
\setlength\tabcolsep{2pt}
\begin{tabular}{clclclcl}
(a) &
\includegraphics[width=1.33in,align=t]{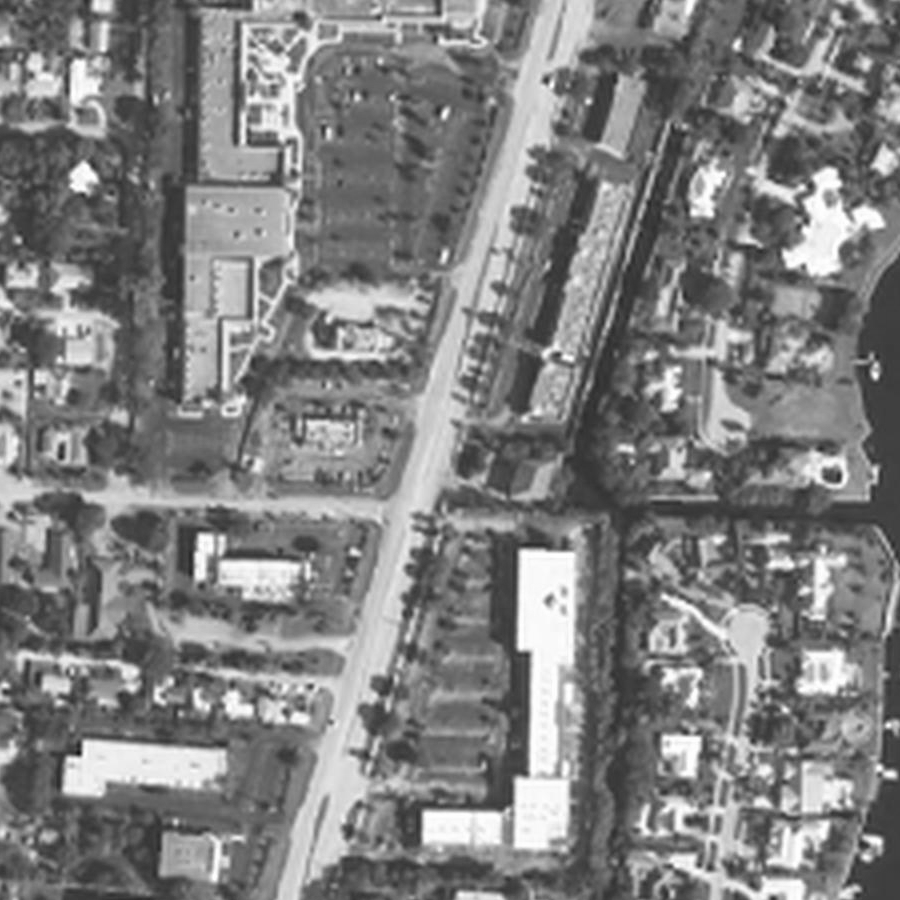} &
(b) &
\includegraphics[width=1.33in,align=t]{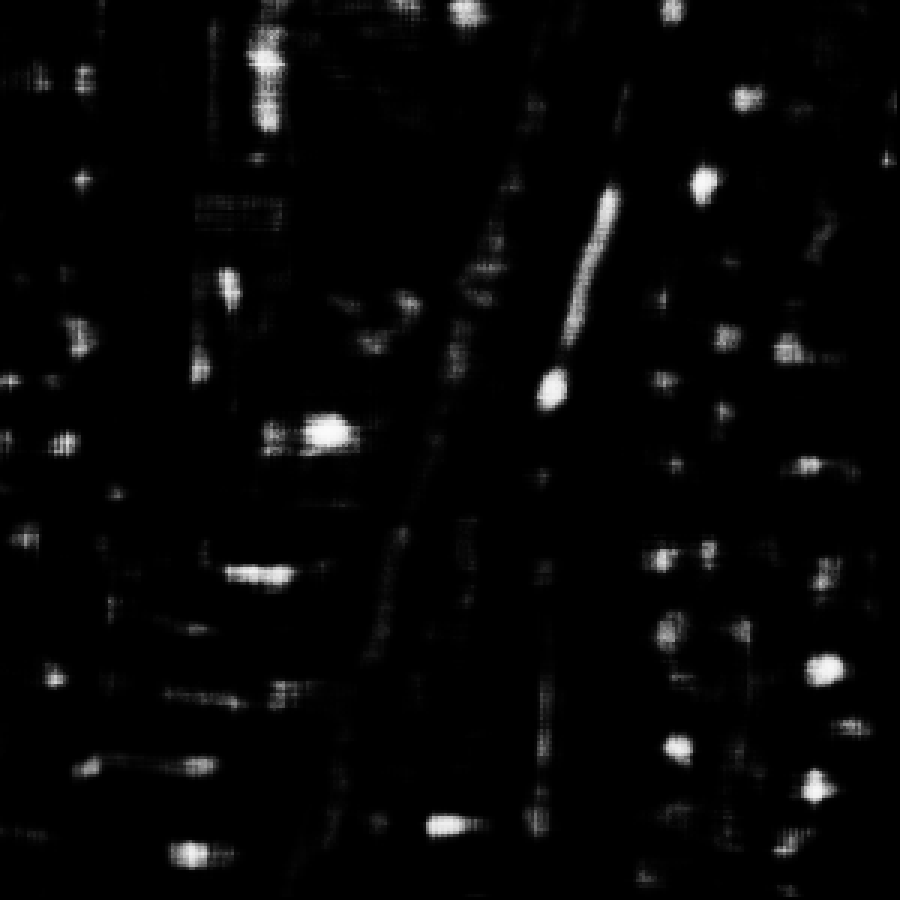} &
(c) &
\includegraphics[width=1.33in,align=t]{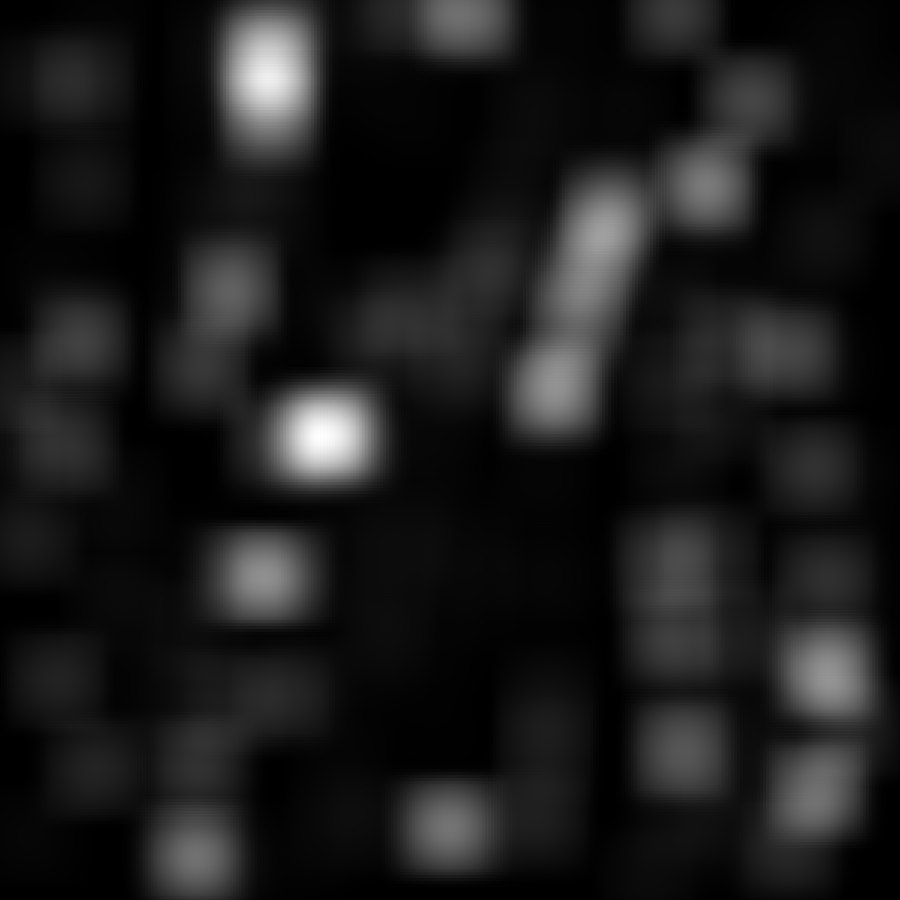} &
(d) &
\includegraphics[width=1.33in,align=t]{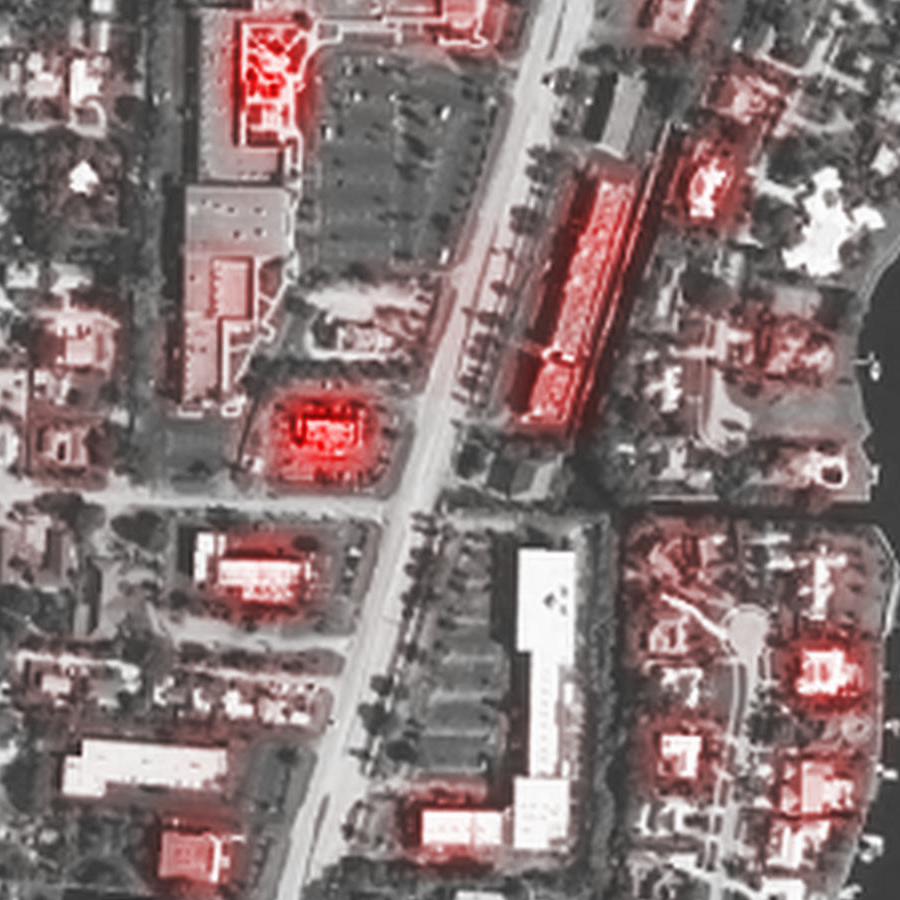} \\
\end{tabular}
\caption{(a) satellite image; (b) pixel-by-pixel classifier outputs; (c)
smoothed classifier outputs; (d) incorporating classifier outputs into image
via soft highlighting.}
\label{fig:highlighting_seq}
\end{figure}

Given an image like that in Figure~\ref{fig:highlighting_seq}a, the
classifier is presented with each $100\times100$ pixel image patch and 
the target probability obtained from the classifier is associated with the
center pixel of the patch.  Figure~\ref{fig:highlighting_seq}b shows the 
probability by pixel. (Only complete patches are classified; thus, there is
a 50-pixel border around the image in which targets will never be detected.)   % RTK:
%{\color{red} [RON, How do you deal with the fact that 100 is an even number   % integer division, so the output is one
%and thus the 'center' pixel isn't exactly in the center? Where is it?]}       % pixel shifted which is inconsequential
Because a target has a spatial extent that is roughly the size of a            % because of the smoothing and single stepped
$100\times100$ patch, the classifier outputs are smoothed using a Gaussian     % sliding window
filter with $\sigma = 50$ pixels, resulting in an activation map such as 
that in Figure~\ref{fig:highlighting_seq}c. 

The activation map is used to determine highlights.  For
soft highlights, we express each pixel in the original image using
hue-lightness-saturation coordinates, and transform the pixel by setting its
saturation level to the value of the corresponding pixel in the activation map
(0 is minimum saturation, 1 is maximum).  The lightness value is obtained from
the original image, and the hue in all our experiments is fixed to be red. This
manner of highlighting preserves information in the original greyscale image
while enhancing the saliency of highlighted locations.  For hard highlights, we
developed a heuristic scheme that: (1) finds the image location of the largest
value in the activation map, (2) draws in the original image a $100\times100$
pixel red square centered on the location, (3) resets the activation map
in a $150\times150$ pixel region around the identified location, (4) repeats
the procedure until the largest activation map value is less than a given
threshold, $\theta$.  This greedy procedure ensures that hard highlights will
not overlap. We selected $\theta=0.5$, which produced between 2 and 15
highlights per image, with a mean and median of 6. See the right panel of
Figure~\ref{fig:hardversussoft} for an illustration of hard highlighting.

\section{Experiments with soft highlighting in satellite imagery}
\subsection{Experiment 5: Target localization with feedback}

In Experiment 5, participants searched images that each contained exactly one
target and were instructed to click where they believed the target was 
located. Following an incorrect click, a buzzer sounded and the trial continued
until the target was correctly located.  Three highlighting conditions were
studied: soft, hard, and a control with no highlighting.

\subsubsection{Stimuli and design}
Stimuli were drawn from a set of 245 target-present images, each
$1000\times1000$ pixels. The target placement was random, with the constraint
that no portion of the target appear in the outer 50 pixel
perimeter of the image. The images were drawn from a test set not used for
training our classifier; thus, the highlights provided by the classifier
reflect the quality of highlighting one would anticipate in actual usage. 
After the image was classified, the outer 50-pixel border was stripped off
and the image was rescaled by a factor of 2/3 to $600\times600$ pixels
to ensure that it would fit entirely within a participant's browser window.

Each participant viewed two blocks of 17 trials, with the first two trials
in a block considered to be practice to familiarize with the task and displays.
All stimuli in a block were highlighted in the same manner (soft, hard, or
control). Counterbalancing was performed across a set of 6 participants, with
each participant seeing the same sequence of images but with a different
pairing of highlighting conditions in the two blocks. (The 6 pairings are:
soft-hard, hard-soft, soft-control, control-soft, hard-control, and 
control-hard.) This counterbalancing ensured that each image occurred
at the same position in a block in each condition, and that each
highlighting condition occurred equally often in blocks 1 and 2.

\subsubsection{Participants}
Participants were restricted to being from North America in order to ensure
familiarity with the typical setting and lay out of fast food restaurants.
Of 289 participants who enrolled in the experiment, 84 (14 groups of 6) 
completed it successfully. (Fourteen groups were chosen in order to use
each image roughly twice across the pool.) Of the 205 participants who did 
not complete, 100 of them quit voluntarily, and 105 were rejected for 
defocusing their browser window. They were instructed not to defocus during
a trial because multitasking corrupts our measure of latency. 

\subsubsection{Procedure}
Prior to the start of the experiment, all images were preloaded on the
participant's machine to ensure smooth pacing of the experiment. In the initial
instructions, participants were told the name of the restaurant chain
(McDonald's) and were shown 20 small example images of the restaurant
(Figure~\ref{fig:representative_examples}) to familiarize themselves with the
task. They were asked not to multitask as Turk workers are apt to do.
At the start of each block, they were shown a representative satellite image
for the highlighting condition, and they were given instructions specific to
the highlighting condition. For soft highlighting, they were told, ``You will
now see a series of 17 black and white images that have been shaded where the
computer believes a restaurant may be;'' and for hard highlighting, ``You will
now see a series of 17 black and white images that have red boxes where the
computer believes a restaurant may be. The actual restaurant may be outside any
of the boxes.'' In both conditions, they were instructed, ``The computer is not
perfect but can provide assistance in locating the restaurant.''

Participants continued to click on the image until the target was detected.  We
considered a click to be on the target if it landed within 30 pixels of
recorded target location in both the horizontal and vertical directions (which
is within 45 pixels of the unscaled image---roughly the full extent of the
restaurant). After each trial, a `next' button appeared to continue to
the following trial.

\subsubsection{Results}
\begin{figure}
\centering
\setlength\tabcolsep{2pt}
\begin{tabular}{llll}
(a) &
\includegraphics[width=2.5in,align=t]{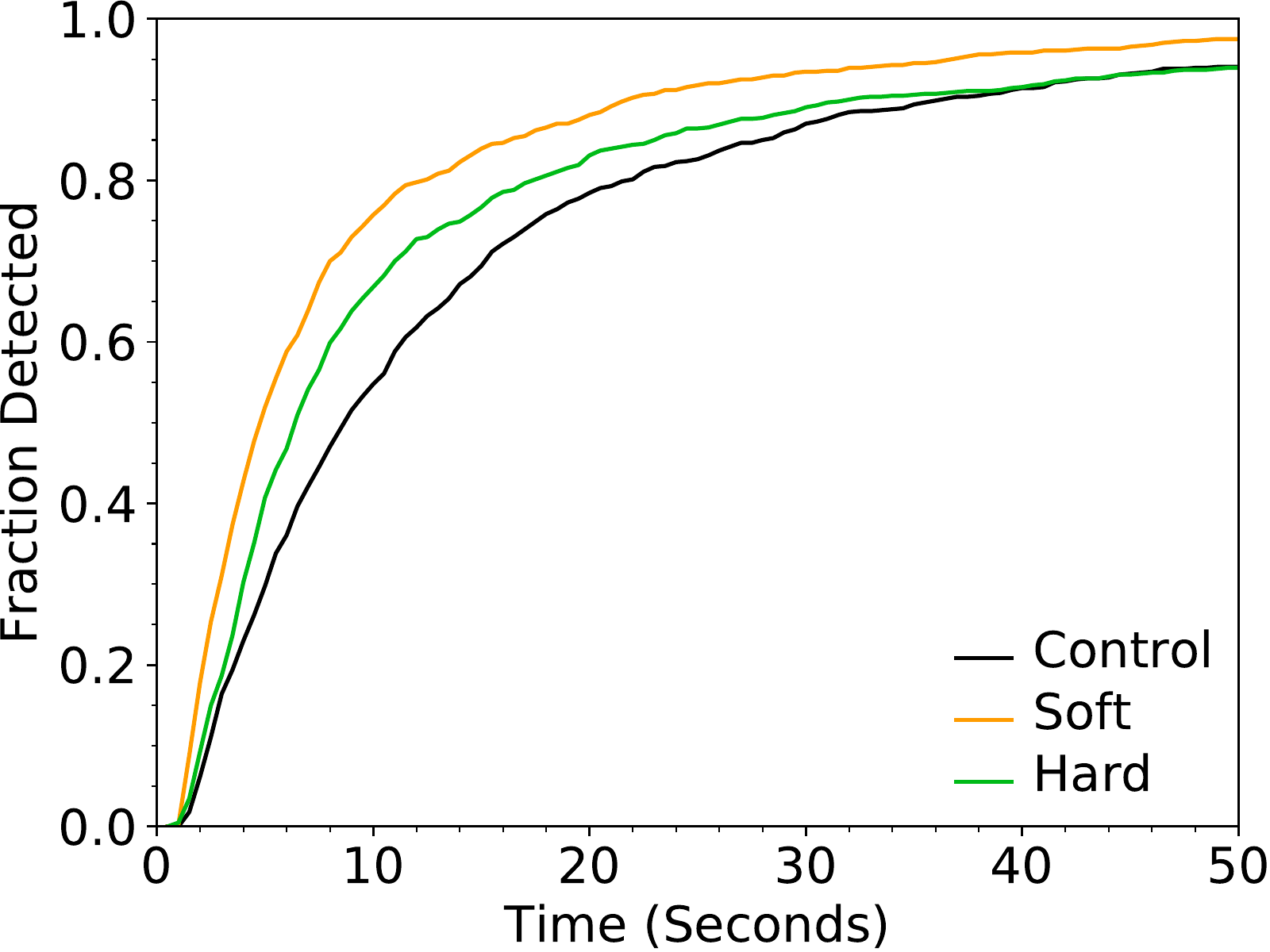} &
(b) &
\includegraphics[width=2.5in,align=t]{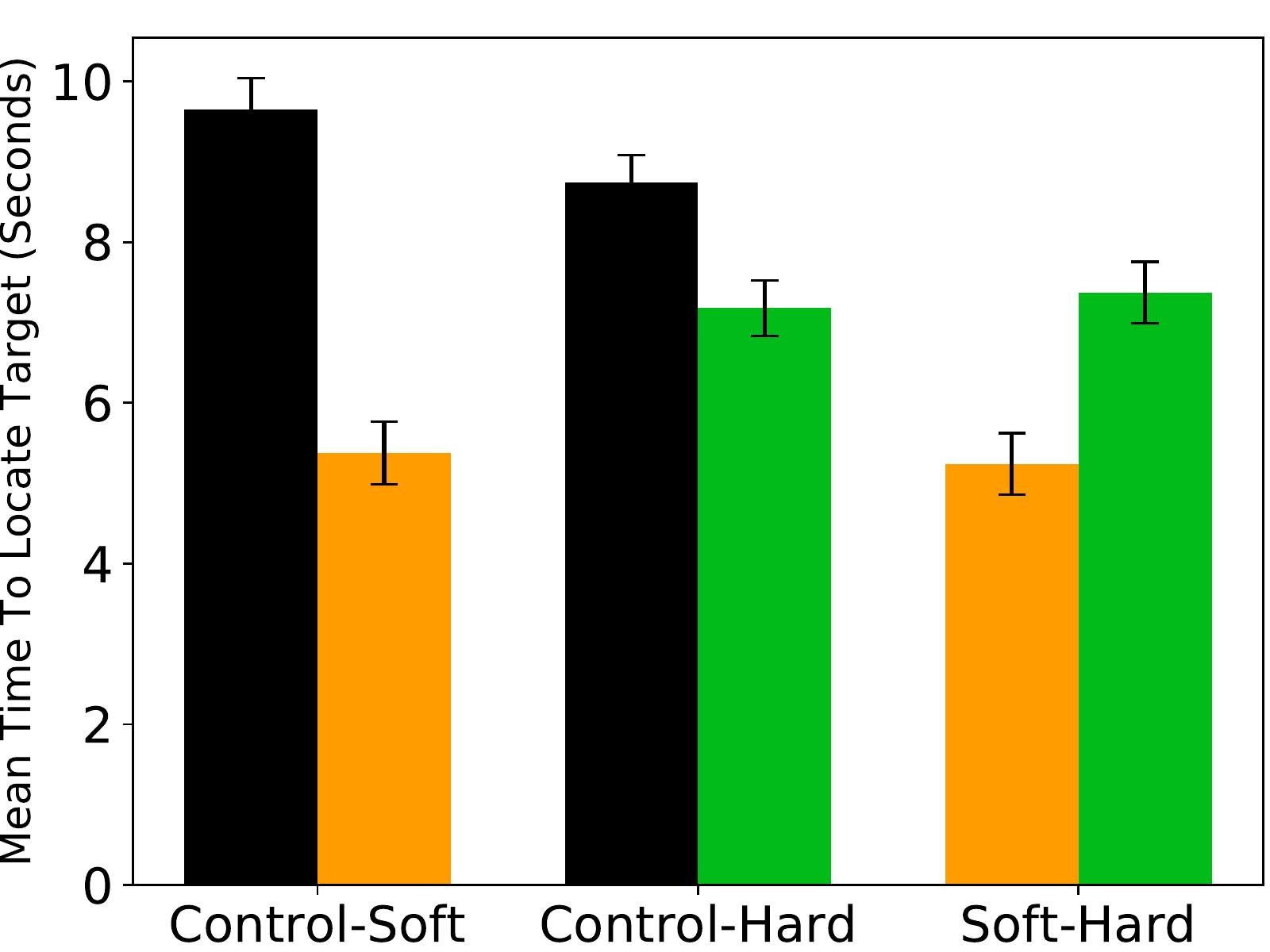} 
\end{tabular}
\caption{Experiment 5: (a) Fraction of targets detected by time and condition
(b) Within-participant comparison of mean search time for pairs of 
highlighting conditions.
%Error bars are calculated using a between-subject
%variability correction \cite{Masson2003}.
}
\label{fig:E5_results}
\end{figure}

Figure~\ref{fig:E5_results}a plots the fraction of targets found by time
within a trial for each condition. Better performance is indicated by a steeper
curve and/or a higher asymptote.  Soft highlighting appears to outperform hard
which in turn outperforms the control condition.  This impression is supported
by within-participant comparisons for pairs of conditions using the
median target-detection latency for each participant and each
condition. The mean across participants are depicted in
Figure~\ref{fig:E5_results}b. Starting with the group on the left, tested
with soft highlighting and the control condition, soft is faster than the
control (9.7 versus 5.4 sec), significant by a paired $t$-test and by the
nonparametric two-sided Wilcoxon signed-rank test ($t(27)=5.40, p<.001$;
$w(27)=1132, p<.001$). Soft highlighting beats hard highlighting as well (5.2
versus 7.4 sec; $t(27)=2.73, p=.011$; $w(27)=539, p<.001$). These analyses
excluded the first two trials of a block as practice; including the two trials
does not affect the outcome of the statistical tests.

%control-soft, w=       1132.0000, p=   4.4197099e-08
%control-hard, w=       1012.0000, p=   2.1938336e-05
%soft-hard,    w=       539.00000, p=   2.1938336e-05

The mean number of nontarget clicks per trial is 7.6, 6.1, and 3.6
in the control, hard, and soft conditions, respectively.
Unpaired $t$-tests indicate that soft highlighting leads to
significantly fewer clicks than either hard 
($t(110)=2.75, p=.007, \mathrm{Cohen's~} d = 0.53$) or control
($t(110)=3.88, p<.001, \mathrm{Cohen's~} d = 0.77$), though the 
difference between hard and control is not reliable
($t(110)=1.19, p=.24, \mathrm{Cohen's~} d = 0.23$).

We examined the relationship between an individual's target-detection latency
and the classifier output probability at the target location.  We found the
strongest correlation with soft ($\rho_{\mathrm{soft}}=-0.43$), then hard
($\rho_{\mathrm{hard}}=-0.25$), then the control
($\rho_{\mathrm{control}}=-0.09$), suggesting that soft highlighting is
effective in communicating classifier confidence.

%are sensitive
%to the gradation of soft highlights and that 
% allows individuals to align their search with classifier confidence

\subsection{Experiment 6: Target detection with feedback}
In Experiment 5, trials terminated only when a target was correctly detected;
therefore, false negative trials could not occur. This design avoided a key
potential pitfall of highlighting: an increase in the probability of missing a
target that is not highlighted (see Section~\ref{sec:no_free_lunch}).
Experiment 6 was therefore designed to allow for the possibility of false
negative responses. Both target present and target absent trials were included,
and participants were allowed to terminate a trial without finding a target.

\subsubsection{Stimuli and design}
Target-present images were re-used from Experiment 5. Target-absent images were
selected as explained in Section~\ref{sec:classifier}.  The experiment
consisted of 36 trials arranged in 4 blocks of 9. Each block included three
images for each highlighting condition (soft, hard, control).  Targets were
present in two of the three images. The order of the 9 trials within a block
was randomized.  We counterbalanced the assignment of images to conditions 
using a Latin square design and testing participants in groups of 3. For each 
set of 3, a unique set of images was used.

\subsubsection{Participants}
Of the 155 participants who enrolled in the experiment, 90 completed 
successfully (30 groups of 3). Of the remaining 65, 43 were rejected for 
changing window focus and the remainder quit voluntarily. Turk workers were 
excluded if outside the US or if they had done previous studies. 

\subsubsection{Procedure}
Because highlighting conditions were intermixed, initial instructions explained
both soft and hard highlighting. Each trial proceeded as in Experiment~5,
except that the screen included a `No restaurant present' button, which allowed
participants to terminate a trial without finding the target. As in Experiment
5, when a location was clicked, the trial terminated if the location contained
a target; otherwise, a buzzer sounded and the trial continued. Following each
trial, a `next' button appeared to continue, and between blocks the
instructions were re-displayed to break up the monotony of the trial sequence.

\subsubsection{Results}
\begin{figure}
\centering
\setlength\tabcolsep{2pt}
\begin{tabular}{llll}
(a) &
\includegraphics[width=2.5in,align=t]{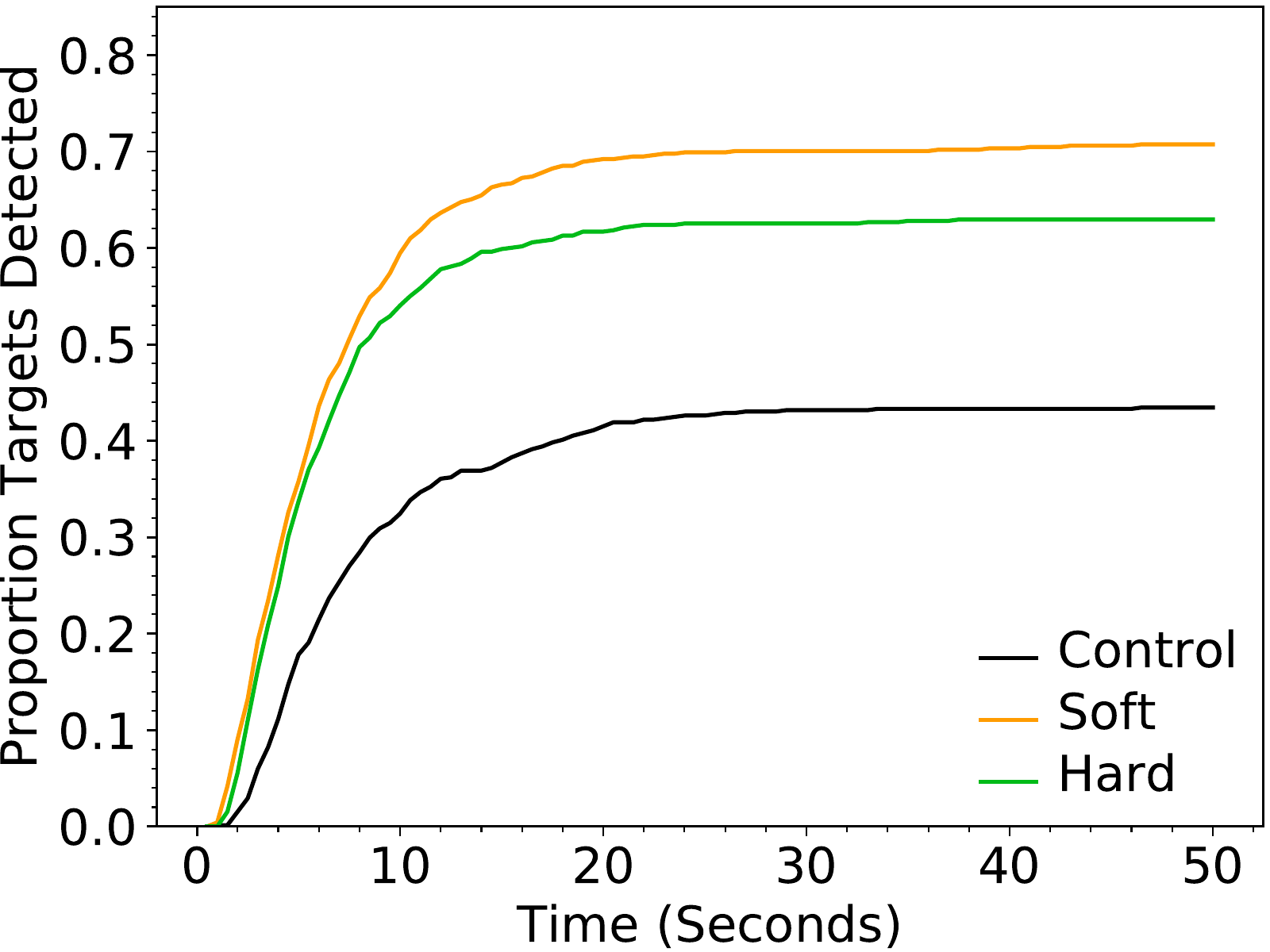} &
(b) &
\includegraphics[width=2.5in,align=t]{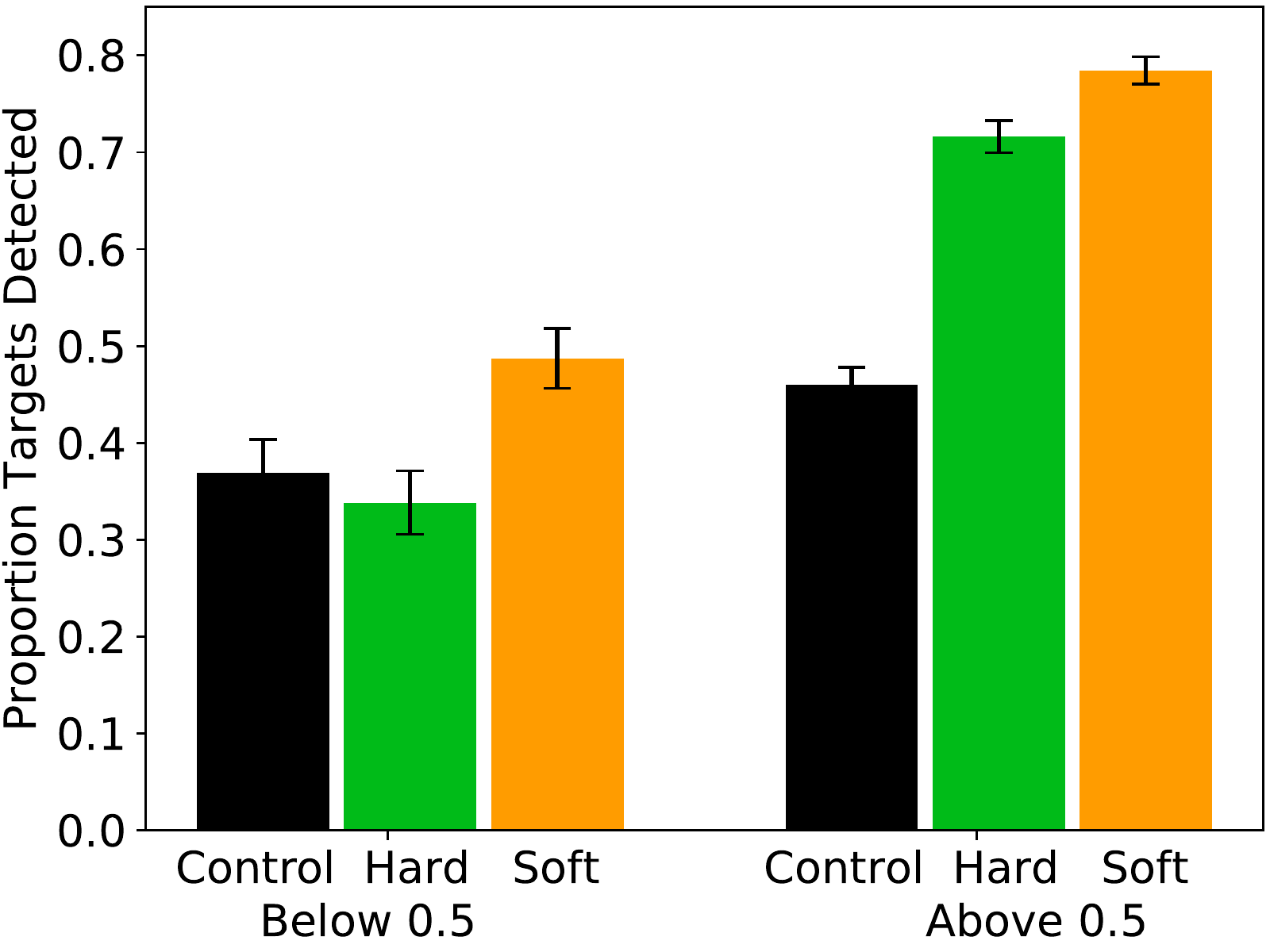} \\
\end{tabular}
\caption{Experiment 6, target-present trials: (a) Fraction of targets 
detected by time and condition; (b) target-detection probability by 
classifier output and condition. 
Error bars are calculated using a between-participant
variability correction \cite{Masson2003}.
}
\label{fig:E6_results}
\end{figure}

Figure~\ref{fig:E6_results}a shows the fraction of targets detected by
highlighting condition as a function of time from stimulus onset. The
curves in this Figure do not asymptote at 1.0 because participants may
give up and miss a target. The asymptotic detection rate for soft, hard,
and control are 70.7\%, 62.8\%, and 43.9\%, respectively. All pairwise
differences are highly reliable by a nonparametric two-sided Wilcoxon 
signed-rank test (soft-control, $w(89)=10618, p<.001$; hard-control, 
$w(89)=9982, p<.001$; soft-hard, $w(89)=7153, p=.004$), as well as by
paired $t$ tests that make stronger assumptions about the data distributions.

On target-absent trials, mean time to terminate is longer for soft (11.3
sec) than for hard (9.6 sec) or control (9.4 sec).  Performing a $t$-test on
the log-transformed times, both the soft-hard ($t(89)=2.94, p=.004$) and
soft-control ($t(89)=3.47, p<.001$) differences are reliable. It's not entirely
clear how to weigh this additional time spent on target-absent trials against
the improvement in both speed and accuracy on target present trials
(Figure~\ref{fig:E6_results}a).

%Control-soft: 9.3622 vs 11.3497, (t(89)=-3.4773, p=0.0008)
%Control-hard: 9.3622 vs 9.6168, (t(89)=-0.5762, p=0.5660)
%Soft-hard   : 11.3497 vs 9.6168, (t(89)=2.9436, p=0.0041)

A critical question this experiment was designed to address is whether
non-highlighted targets are missed more often when highlighting is present.
Figure~\ref{fig:E6_results}b suggests that soft highlighting may potentially be
a solution to the `no free lunch' dilemma. We have taken all target-present
trials and broken them up by whether the classifier obtained a confidence level
above or below 0.5---the threshold we use for hard highlighting.\footnote{We
did not control for classifier output level in each participant's trial
sequence.  Consequently, some participants had no trials in certain conditions
with below 0.5 confidence levels. Our classifier has high confidence for
targets on 76\% of all trials.} For high-confidence classifier outputs, soft
highlighting produces more detections than hard ($t(178)=3.12, p=.002$), and
both soft and hard highlighting beat the control condition ($t(178)=13.99,
p<.0001$, $t(178)=10.26, p<.0001$). For low-confidence classifier outputs, soft
highlighting beats both hard and the control ($t(156)=3.06, p=.003$,
$t(156)=2.38, p=.018$). We do not observe a cost for hard highlighting relative
to the control ($t(156)=.61, p=.54$), which we expected to see based on the
existing literature suggesting that hard highlighting can mask non-highlighted
targets.  Nonetheless, the advantage of soft highlighting over hard in
Figure~\ref{fig:E6_results}a seems to hold regardless of the classifier output.

%Above:
%
%    control-hard:  0.45979 +/- 0.01840  vs  0.71613 +/- 0.01669  (t(178)=10.26239, p=0.00000)
%    control-soft:  0.45979 +/- 0.01840  vs  0.78448 +/- 0.01395  (t(178)=13.98585, p=0.00000)
%    soft-hard   :  0.78448 +/- 0.01395  vs  0.71613 +/- 0.01669  (t(178)=3.12460, p=0.00208)
%
%
%Below:
%
%    control-hard:  0.36941 +/- 0.03395  vs  0.33840 +/- 0.03289  (t(156)=0.61080, p=0.54222)
%    control-soft:  0.36941 +/- 0.03395  vs  0.48713 +/- 0.03100  (t(156)=2.38378, p=0.01834)
%    soft-hard   :  0.48713 +/- 0.03100  vs  0.33840 +/- 0.03289  (t(156)=3.06381, p=0.00258)

\subsection{Experiment 7: Target detection without feedback} 

In Experiment 6, false-positive clicks received immediate feedback.
In our final experiment, we removed all feedback during a trial, allowing
participants to select one location if they believed a target was present or to
indicate that no target was present.  Feedback was provided after a
response was submitted. (Without any feedback, we were concerned that
participants would fail to learn the task and might also lose motivation.)
With the possibility of both false-positive (FP) and false-negative (FN) 
responses, this experiment allows us to analyze the effect of highlighting 
on target \textit{discriminability}, using measures such as $\dprime$ and 
AUC.

\subsubsection{Stimuli and design}
The stimuli and design were identical to Experiment 6. Participants
were presented with 36 images in four blocks of 9 trials.

\subsubsection{Participants}
Of the 130 participants who enrolled in Experiment 8, 90 completed
successfully. Of the 40 who did not complete, 16 left voluntarily and 24 were
rejected for de-focusing their browser windows during a trial.  Turk workers
were excluded if outside the US or if they had done previous studies.

\subsubsection{Procedure}
Once a trial began, participants were allowed to click anywhere in the image to
indicate a target. Clicking displayed a cross hair (plus sign) at the clicked
location.  If another location was clicked, the cross hair moved. Thus,
at most one cross hair was present on the screen. No feedback was provided
as the participant clicked. Below the image were two buttons. Clicking the
button labeled `no restaurant present' removed a cross hair from the display
if one was present. Clicking the button labeled `submit response' terminated
the trial and recorded the location of the cross hair if one was present.
After each trial, participants received feedback indicating whether
their response was correct or not. A reduced-scale version of the image was 
shown with the location of the actual target, if any. Additionally,
a pleasant ding or an annoying buzzing sound was played depending on 
response correctness. Feedback remained on the screen until participants
clicked a `next' button to continue the experiment.

\subsubsection{Results}
\begin{figure}
\centering
%\setlength\tabcolsep{0pt}
%\begin{tabular}{llllll}
(a) 
%\begin{minipage}[b]{1.87in}
\includegraphics[width=1.87in,align=c]{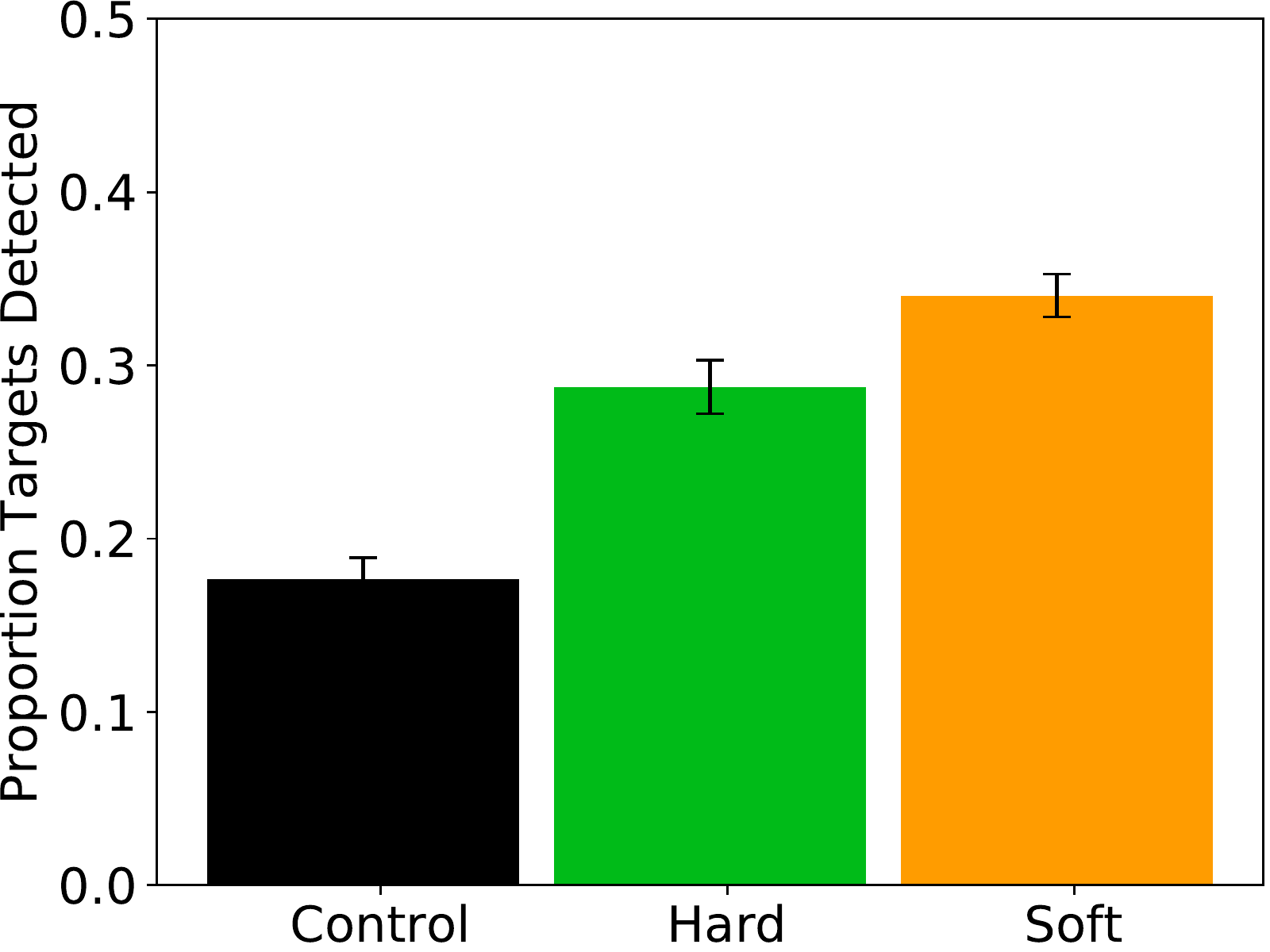} 
%\end{minipage}
(b) 
%\begin{minipage}[b]{1.87in}
\includegraphics[width=1.87in,align=c]{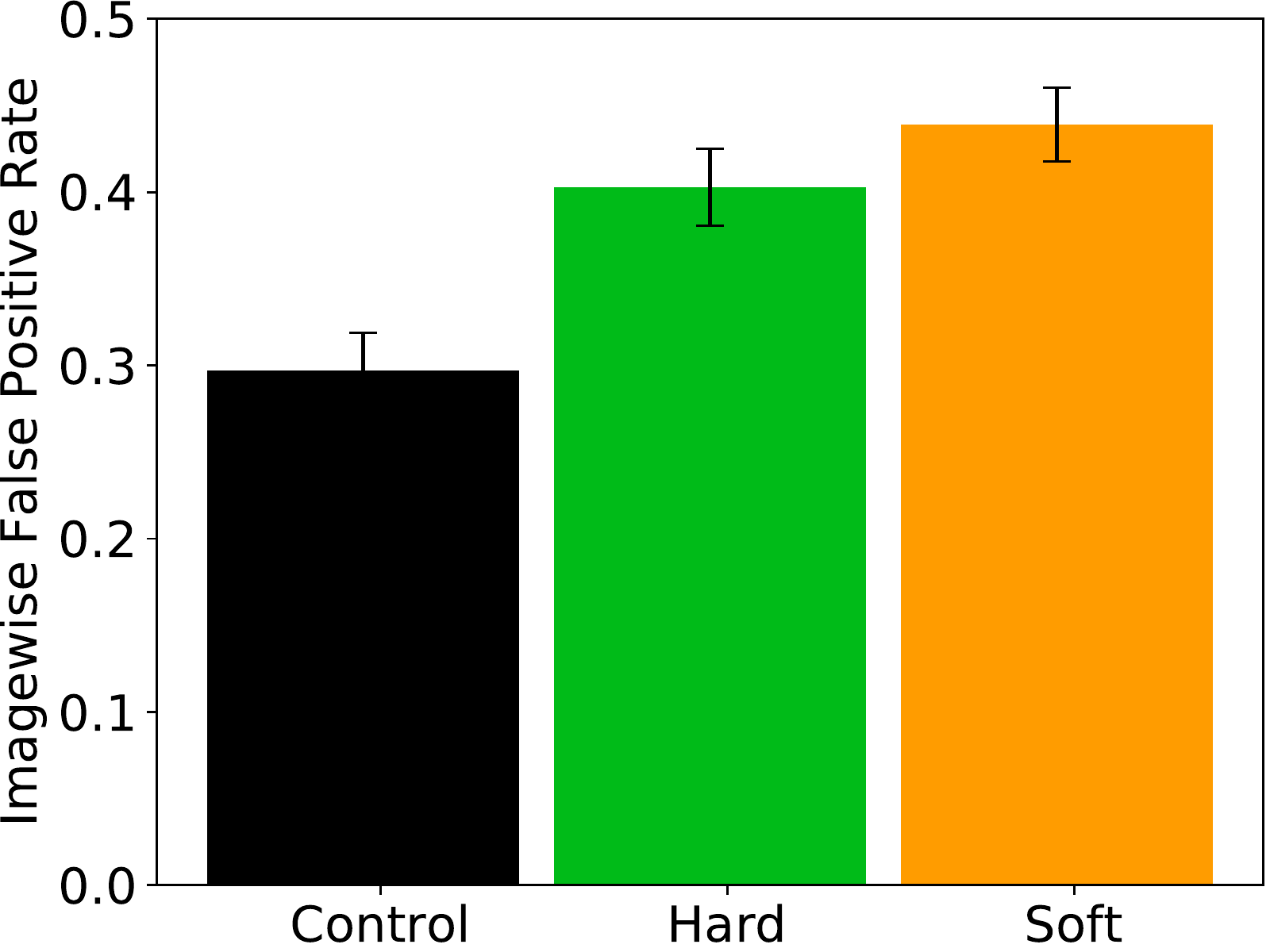} 
%\end{minipage}
(c) 
\includegraphics[width=2.2in,align=c]{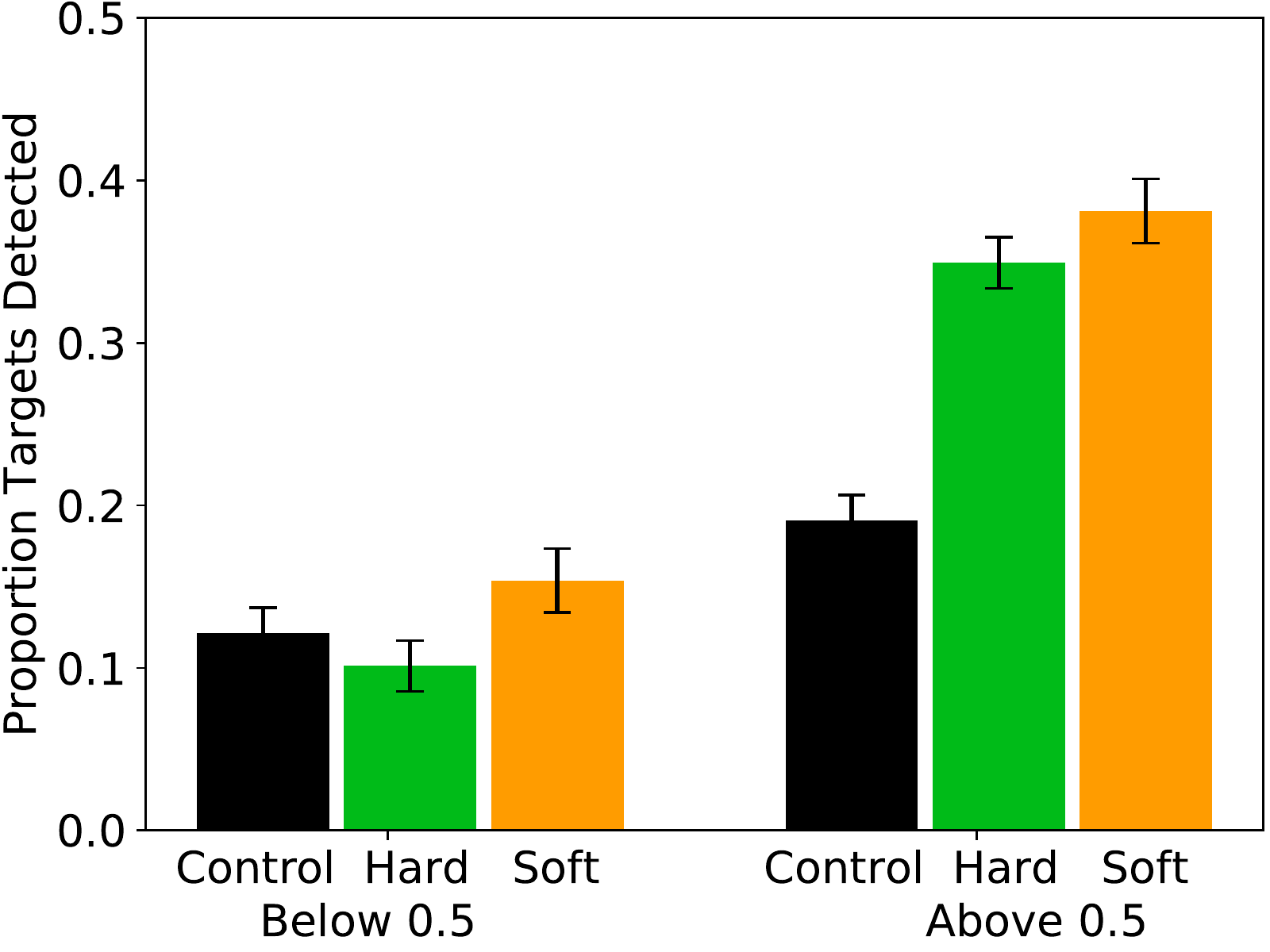} 
%(a) &
%\includegraphics[width=1.87in,align=t]{E7_results_tp.pdf} &
%(b) &
%\includegraphics[width=1.87in,align=t]{E7_results_fp.pdf} &
%(c) &
%\includegraphics[width=2.2in,align=t]{E7_detection_given_highlights.pdf} 
%\end{tabular}
\caption{Experiment 7: (a) proportion of target-present trials in which target
is detected (TP rate), (b) proportion of target-absent trials in which some 
location is misidentified as a target (image-wise FP rate).
(c) proportion of targets detected when classifier
response to target was below or above 0.5.
}
\label{fig:E7_results}
\end{figure}

%\label{fig:E7_detection_given_highlights}

Using target-present trials, we compute the proportion targets detected
(TP rate) by condition (Figure~\ref{fig:E7_results}a). Fewer targets are 
detected with the no-highlight control than hard highlighting (0.18 versus 
0.29, $t(89)=5.64,
p<.0001$) or soft highlighting (0.18 versus 0.34, $t(89)=6.47, p<.0001$).
Fewer targets are detected with hard highlighting than soft (0.34 versus 0.29,
$t(89)=2.10, p=.039$).  As in Experiment 6 (Figure~\ref{fig:E6_results}a,b),
lower asymptotic detection rates in Experiment 7 go hand in hand with slower
detection (not shown), and the benefit of soft highlighting is observed
regardless of classifier confidence (Figure~\ref{fig:E7_results}c).

%TP:
%  Control-soft: 0.1764 vs 0.3403 (t(89)=-6.46647, p=0.000000)
%  Control-hard: 0.1764 vs 0.2875 (t(89)=-5.64103, p=0.000000)
%  Soft-hard   : 0.3403 vs 0.2875 (t(89)=2.09949, p=0.038604)
%
%FP:
%  Control-soft: 0.2972 vs 0.4389 (t(89)=-3.67835, p=0.000401)
%  Control-hard: 0.2972 vs 0.4028 (t(89)=-2.86509, p=0.005201)
%  Soft-hard   : 0.4389 vs 0.4028 (t(89)=0.95014, p=0.344614)

Using target-absent trials, we compute the proportion of trials that
\textit{some} location is selected as the target, which we refer to as the
\textit{image-wise FP rate} (Figure~\ref{fig:E7_results}b).
Consistent with the literature discussed in Section~\ref{sec:no_free_lunch},
highlighting---both soft and hard---leads to more locations erroneously
identified as targets (soft versus control, $t(89)=3.68, p<.001$; hard versus
control $t(89)=2.87, p=.005$); however, the difference between soft and hard
is not significant ($t(89)=0.95, p=.34$).

%{\color{red} RON: your w value for soft-versus-hard is wrong. See Figure 5.17
%of your thesis. If w=7338 is significant, w=8327 must be too. Please go back
%and check ALL statistics throughout this article. This type of error will get
%our paper rejected.}

Although it is encouraging that the TP rate is higher for soft than hard
highlighting and that the FP rates do not differ significantly, one would like
to compare conditions using a measure of discriminability like $\dprime$.
However, in order to do so, we must address an issue concerning the computation
of FP rates.  A false positive occurs when a participant mistakenly labels a
patch as a target.  Each target-absent image contains many patches, and if
\textit{any} of these patches is seen as a target, it will contribute to the
image-wise FP rate. Assuming that the scoring of one patch is independent of
another, we can recover the patch-wise FP rate from the image-wise rate:
%\begin{equation*} %\textit{FP}_{\textrm{image}} = %1 -
%\textit{TN}_{\textrm{image}}\\ %= 1 - \textit{TN}_{\textrm{patch}}^P ,
%\end{equation*} 
\begin{equation*} 
\textit{FP}_{\textrm{patch}} = 1 - (1 - \textit{FP}_\textrm{image})^\frac{1}{P} , 
\end{equation*} 
where $P$ is the effective number of patches.  We use
$\textit{FP}_\textrm{patch}$ for estimating $\dprime$. Although the choice of
$P$ has a scaling effect on $\dprime$, $P$ does not influence any comparisons
we make across conditions (i.e., relative values of $\dprime$).  We used
$P=100$ to reflect the fact that targets lie in a roughly $100\times100$ pixel
region of our $1000\times1000$ pixel images, yielding $10\times10$
nonoverlapping patches.

Table~\ref{tbl:e7_overall} (middle column) shows per-condition $\dprime$ 
computed by pooling data across participants to obtain patch-wise TP and FP rates for the population.
Soft highlighting outperforms hard, and both outperform the control.  It is
informative to compute $\dprime$ for the classifier alone. In order to do so,
we must pick a classifier-output threshold for judging a target as being
present.  We picked two different thresholds in order to match either the FP or
the TP rate of our human population. Although these two thresholds yield
slightly different $\dprime$ values, both are only slightly better than humans
in the control condition, and both are worse than humans in the highlighting
conditions.  Thus, the hybrid human-classifier judgments, obtained via soft or
hard highlighting, are superior either to human judgments alone (the control
condition) or classifier judgments alone.

\begin{table}
\tbl{Human and classifier discriminability scores. First column obtained by
pooling human data across participants, and using $P=100$ and $Q=0$. Second 
column obtained by computing $\dprime$ for each participant and then 
averaging, using $P=100$ and $Q = 1$. Standard deviations of the mean
are shown in parentheses, corrected for inter-participant variability 
\cite{Masson2003}.}{
\begin{tabularx}{\textwidth}{p{1in} p{1.25in} | l | l l }
\hline
    & &   pooled $\dprime$ & individual $\dprime$ mean & individual $\dprime$ SD \\ \hline
human & control condition   &  1.90 & 2.07 & (0.38) \\
human & hard highlighting   &  2.49 & 2.45 & (0.37) \\
human & soft highlighting       &  2.70 & 2.60 & (0.47) \\
classifier & matched FP rate    &  2.19 & 2.20 & \\
classifier & matched TP rate    &  1.97 & 2.08 & \\
\end{tabularx}}
\label{tbl:e7_overall}
\end{table}
%\begin{table}
%\tbl{Human and classifier $d^{\prime}$. Human values obtained by pooling 
%data across participants. All values computed with $P=100$}{
%\begin{tabular}{c c | l}
%    & &   $d^{\prime}$ \\ \hline
%human & control condition   &  1.90 \\
%human & hard highlighting   &  2.49 \\
%human & soft highlighting       &  2.70 \\
%classifier & matched FP rate    &  2.19 \\
%classifier & matched TP rate    &  1.97 \\
%\end{tabular}}
%\label{tbl:e7_overall}
%\end{table}

%Overall d' (simply counting TP, FP, TN, FN by condition) (Q=0.0, locations=100):
%
%control = 1.8991
%hard    = 2.4875
%soft    = 2.7018
%
%match human TN rate, classifier d' = 2.1910 (Q=0.00, locations=100, ctn=0.9897, htn=0.9898, theta=0.9471)
%match human TP rate, classifier d' = 1.9652 (Q=0.00, locations=100, ctp=0.3381, htp=0.3369, theta=0.9548)

These population-wide $\dprime$ values say nothing about
how reliably \textit{individuals} discriminate targets from nontargets across
conditions.
To do so, we must calculate $\dprime$ for each individual, but estimates
are quite noisy due to the small number of trials in each condition:
per participant and condition, there were only 4 target-absent trials to
compute a FP rate, and only 8 target-present trials to compute a TP rate.
Because the TP and FP rates could easily be 0.0 or
1.0, yielding ill-defined $\dprime$ values, we performed smoothing based
on the Bayesian notion of imaginary-count priors. Essentially, we
assume a constant $Q$ that regularizes the computed rates:
\begin{equation*}
\textit{TP} = \frac{C+Q}{8+2Q} \textrm{~~and~~} \textit{FP} = \frac{E+Q}{4+2Q} ,
\end{equation*}
where $C$ is the number of targets correctly detected in target-present images,
and $E$ is the number of trials in which a target is erroneously detected in
target-absent images. $Q$ represents the strength of a prior belief that the TP
and FP rates are 0.5, with a larger $Q$ smoothing observations away from the
extreme rates of 0.0 and 1.0. As $Q\to\infty$, the observations matter less and
all values approach 0.5 in the limit. Operationally, smoothing assumes $Q$
additional trials in which a target is detected and $Q$ additional trials in
which no target is detected.

We arbitrarily picked $Q=1$, which limits TP rates to lie in $[0.1, 0.9]$ and
FP rates to lie in $[0.17, 0.83]$. Table~\ref{tbl:e7_overall} (right column) presents $\dprime$
values for the three experimental conditions and the classifier, computed with
$Q$ smoothing. The Table also shows inter-participant variability.
Soft and hard highlighting both
lead to larger $\dprime$ than the control (soft: $t(89)=6.13, p < .001$;
hard: $t(89)=6.65, p<.001$), and soft highlighting is marginally superior to
hard ($t(89)=1.93, p=.057$).  Effect sizes, as measured by Cohen's $d$ are 1.27
(control vs. soft), 1.03 (control vs. hard), and 0.36 (soft vs. hard).

Although the contrasts between the two highlighting conditions and the control
are insensitive to $Q$, the critical contrast between soft and hard is.
As $Q$ increases, the effect size becomes larger and the reliability of
the $t$ test increases. Rather than picking $Q$ arbitrarily, we used
statistical tests of non-normality to determine the smallest $Q$ that yields a
data distribution that did not deviate systematically from a Gaussian.
Using the Jarque-Bera or D'Agostino test of normality to pick $Q$, we obtain 
still-marginal soft versus hard contrast ($t(89)=2.03, p=.045$).

\subsection{Conclusions from experiments with satellite imagery}

Soft highlighting proved robustly superior to hard highlighting on satellite 
imagery search in Experiments 5 and 6, which prevented false positive
responses via intra-trial feedback.  In Experiment 7, when no intra-trial
feedback was provided, false positives did occur, but when true
and false positives are combined to obtain a $\dprime$ discriminability 
measure, soft highlighting is marginally better than hard highlighting.
In retrospect, our Experiment 7 may not have had adequate power due to the
small number of trials per participant and condition, and we expect that
a larger replication would obtain a more robust contrast.

Across Experiment 5-7, highlighting conditions are superior to the control
condition with no highlighting. Given that our participants were relative
novices at the particular search task, it seems sensible that highlights
would facilitate performance. It remains an interesting open question
how much domain experts would benefit from highlights, and the relative
benefit that soft and hard highlights would provide.

\section{General Discussion}

To boost human performance on difficult image analysis tasks,
techniques have been developed to augment images with boxes or arrows to
indicate locations that
warrant close scrutiny. However, this sort of all-or-none or {\em hard}
highlighting has not proven robustly effective.  Our findings give us optimism
to suppose that graded or {\em soft} highlighting may turn out to be more
beneficial. In both tasks where individuals are domain experts (handprinted
digit search, Experiments 1-4) and where individuals are relative
novices (satellite-imagery search, Experiments 5-7), we have shown a consistent
advantage for soft over hard highlighting.

A very recent manuscript by \citeN{CunninghamDrewWolfe2016} also argues for
the superiority of soft over hard highlighting, or---in their
terminology---\textit{analog} over \textit{binary}. In their experiments,
participants search clusters of colored dots for a target defined by the
statistics of its dots. Clusters are highlighted by rendering a ring around the
cluster. Hard highlighting is performed via rings that are present or absent,
soft highlighting via the shading of the rings. Cunningham et al.\ found a
benefit for soft highlighting, consistent with our results.  Their controlled,
artificial stimuli allow for the objective assessment of stimulus signal
strength, facilitating in principle the investigation of how individuals 
combine signals from the stimulus and highlights. In contrast, our experiments 
have real-world validity, utilizing complex stimuli that are familiar and 
naturalistic. 

Another distinction between our work and that of Cunningham et al.\ is the
manner of highlighting. We modulate image lightness (Experiments 1-4) and
saturation (Experiments 5-7) directly, whereas Cunningham et al. superimpose
distinct display markers to guide attention.  We conjecture that the most
effective means of communicating graded signals from a classifier will be those
that leverage the sensitivity of the human attentional system.  The attentional
system excels in integrating bottom-up stimulus saliency with top-down task
guidance. Our approach, which involves modulating lightness and saturation,
provides natural bottom-up cues to guide attention. Of course, other types of
highlights may be equally effective, e.g., contrast or display flicker. With
cuing via saliency, we have shown that individuals are sensitive to the
gradations of soft highlighting, and that even poor quality classifiers can
boost human visual search. 

To develop even more effective techniques for soft highlighting, we suspect
that it will be valuable to conduct a systematic investigation of techniques
to represent classifier confidence scores in images.  In our work, we aimed
for a linear relationship between classifier confidence and attentional
saliency.  However, it may turn out that convex or concave functions will be
more suitable depending on the nature of the classifier, the domain, or
the highlighting manipulation.

Another interesting direction for future research concerns the complementarity
of humans and machines. One expects the greatest synergy when the skills of
humans and machines are orthogonal. For example, our restaurant classifier
systematically inspected local image patches, which likely made it superior to
humans in picking out the granular features of roofs, parking lots, etc.  In
contrast, humans are able to leverage coarser contextual knowledge, such as the
fact that restaurants are often located on major streets and intersections.
Consequently, the human-classifier combination outperformed either agent alone.

One might question the value of human-machine cooperative methods, given
that machine learning methods are likely to advance to a level where they
match or surpass human expertise, and they do not suffer the attentional and 
motivational lapses that confound human experts. Nonetheless, humans are
still necessary in the loop to provide initial training data and to
refine the data set. The refinement process will be particularly critical
and will involve humans working with machines in much the manner we
examined in this work, in order to catch the subtle cases that either
human alone or machine alone is likely to miss.

%Might expect best synergies when human errors and classifier errors are
%orthogonal.  This is the situation where an ensemble would be particularly
%beneficial (vs.  the case where both make the same errors, and therefore one
%cannot catch the other.) We unintentionally achieved this orthogonality via the
%random aspect of the SOCs: a poorly drawn target '2' was no more likely to have
%low confidence as a prototypical target '2'. with the restaurant-search task,
%our hunch is that there was a synergy because the classifier picked up
%complementary information from what individuals use.  people use broader scene
%context (main road, corners,etc.); classifier depends only on shape of
%restaurant and immediate surround.

%what people know: suburban environments in which restaurants typically
%located: main streets, often corners, drive through, parking lot.

\newpage

% Bibliography
\bibliographystyle{ACM-Reference-Format-Journals}
\bibliography{highlighting}
                                % Sample .bib file with references that match those in
                                % the 'Specifications Document (V1.5)' as well containing
                                % 'legacy' bibs and bibs with 'alternate codings'.
                                % Gerry Murray - March 2012

% History dates
%\received{December 2016}{December 2016}{December 2016}

% NO APPENDIX
%\elecappendix

\end{document}